%% file: main.tex
\documentclass[%
reprint,
superscriptaddress,
preprintnumbers,
nofootinbib,
 amsmath,amssymb,
 aps,prd
]{revtex4-1}

\usepackage[dvipsnames]{xcolor}
\usepackage{hyperref}
\usepackage{blindtext}
\usepackage{comment}
\usepackage{enumitem}
\usepackage{gensymb}
\usepackage{todonotes}

\usepackage{graphicx}
\usepackage{dcolumn}
\usepackage{bm}
\usepackage{fontawesome}
\usepackage{chemformula}
\usepackage[utf8]{inputenc}

\definecolor{steelblue}{RGB}{25,25,112}

\hypersetup{colorlinks,linkcolor={OliveGreen},citecolor={Maroon},urlcolor={steelblue}}

\usepackage{colortbl}

\newcommand{\FIRSTAFF}{\affiliation{Gravitation Astroparticle Physics Amsterdam (GRAPPA), Institute for Theoretical Physics Amsterdam and Delta Institute for Theoretical Physics, University of Amsterdam, Science Park 904, 1098 XH Amsterdam, The Netherlands}}
\newcommand{\SECONDAFF}{\affiliation{École Normale Supérieure Paris-Saclay,
61 Avenue du Président Wilson, 94230 Cachan, France}}
\newcommand{\THIRDAFF}{\affiliation{The Oskar Klein Centre, Department of Physics, Stockholm University, AlbaNova, SE-10691 Stockholm, Sweden}}

\begin{document}

\title{Radio Signal of Axion-Photon Conversion in Neutron Stars:\\ A Ray Tracing Analysis}

\author{Mikaël Leroy}
\email{mikael.leroy@ens-paris-saclay.fr}
\FIRSTAFF
\SECONDAFF

\author{Marco Chianese}
\email{m.chianese@uva.nl}
\FIRSTAFF

\author{Thomas D. P. Edwards}
\email{t.d.p.edwards@uva.nl}
\FIRSTAFF
\THIRDAFF

\author{Christoph Weniger}
\email{c.weniger@uva.nl}
\FIRSTAFF

\date{\today}

\begin{abstract}
Axion dark matter can resonantly convert into photons in the magnetospheres of neutron stars (NSs). It has recently been shown that radio observations of nearby NSs can therefore provide a highly sensitive probe of the axion parameter space. Here we extend existing calculations by performing the first three-dimensional computation of the photon flux, taking into account the isotropic phase-space distribution of axions and the structure of the NS magnetosphere. In particular, we study the overall magnitude of the flux and its possible time variation. We find that overall signal strength is robust to our more realistic analysis. In addition, we find that the variance of the signal with respect to the NS rotation is washed out by the additional trajectories in our treatment. Nevertheless, we show that SKA observations towards J0806.4-4123 are sensitive to $g_{a\gamma\gamma}\sim 3\times10^{-13}\mathrm{\,GeV^{-1}}$ at $m_a\sim 7\times10^{-6}\mathrm{\,eV}$, even when accounting for Doppler broadening. Finally, we provide the necessary code to calculate the photon flux for any given NS system \href{https://github.com/mikaelLEROY/AxionNS_RayTracing}{\faGithub}.
\end{abstract}

\pacs{Valid PACS appear here}
\maketitle
\section{Introduction}

The QCD (quantum chromodynamics) axion was first introduced in 1977 by Peccei and Quinn as a solution to the Strong CP problem of the QCD sector~\cite{Peccei:1977hh,Peccei:1977ur,Weinberg:1977ma,Wilczek:1977pj}. Depending on the production mechanisms at play in the early Universe, the QCD axion can behave like cold collisionless matter, therefore allowing it to account for a fraction, or all, of the dark matter (DM) in the Universe~\cite{Preskill:1982cy,Abbott:1982af,Dine:1982ah}. The QCD axion is therefore considered one of the most well-motivated DM candidates to date. The assumption that the Pecci-Quinn (PQ) symmetry is broken after inflation leads to the axion mass of $m_a =  26.2 \pm 3.4~\mu\mathrm{eV}$~\cite{Klaer:2017ond}, though larger uncertainties may arise from the contribution of topological defects~\cite{Gorghetto:2018myk,Kawasaki:2018bzv,Vaquero:2018tib}. If, on the other hand, the PQ symmetry is broken before inflation this mass constraint can be relaxed to give the classical window of $10^{-7}\,\mathrm{eV} \lesssim m_a \lesssim 10^{-4}\,\mathrm{eV}$~\cite{Wilczek:2004cr,Hertzberg:2008wr,Freivogel:2008qc,Visinelli:2009zm,Hamann:2009yf,Hoof:2018ieb}.
 
A large range of observational strategies to directly detect axion particles now exist~\cite{Asztalos:2009yp,Silva-Feaver:2016qhh,TheMADMAXWorkingGroup:2016hpc,Majorovits:2017ppy,Brun:2019lyf,JacksonKimball:2017elr,Anastassopoulos:2017ftl,Zhong:2018rsr,Du:2018uak,Ouellet:2018beu,Shokair:2014rna,Kenany:2016tta,Brubaker:2016ktl,Kahn:2016aff,McAllister:2017lkb,Alesini:2017ifp,Lawson:2019brd} (see Ref.~\cite{Irastorza:2018dyq} for a recent review). Many of these experiments exploit the axion's coupling to electromagnetism $\mathcal{L} = -  (1/4) g_{a\gamma\gamma} F_{\mu\nu} \tilde{F}^{\mu\nu} \, a = g_{a\gamma\gamma} \textbf{E} \cdot \textbf{B} \,a$ to induce axion-photon conversion in the presence of magnetic fields. For the QCD axion, the axion mass and axion-photon coupling strength are proportional, $m_a \propto g_{a\gamma\gamma}$~\cite{Kim:1979if,Shifman:1979if,Zhitnitsky:1980tq,Dine:1981rt}. On the other hand, axion-like particles (ALPs) do not have the same coupling/mass relation and can therefore take on a wider variety of parameter combinations. Although not connected to the strong CP problem, ALPs are a generic prediction from the spontaneous breaking of approximate global symmetries in beyond the standard model physics as well as compactifications of higher dimensions in string theory~\cite{Conlon:2006gv,Grana:2005jc}. ALPs therefore represent a prime target for searches of new physics.

The most sensitive axion DM detector for masses around $2\,\mu\mathrm{eV} \lesssim m_a \lesssim 4\,\mu\mathrm{eV}$ is the ADMX experiment which uses a cold microwave resonator in a strong magnetic field, $\mathcal{O}(1-10)\,\mathrm{T}$, to induce axion-photon conversion~\cite{Asztalos:2009yp,Du:2018uak}. In the future, HAYSTAC~\cite{Zhong:2018rsr} and MADMAX~\cite{TheMADMAXWorkingGroup:2016hpc,Majorovits:2017ppy,Brun:2019lyf} will probe a similar mass range. ABRACADABRA~\cite{Ouellet:2018beu} and DM-radio~\cite{Silva-Feaver:2016qhh} will probe ALPs at lower masses $10^{-10}\,\mathrm{eV} \lesssim m_a \lesssim 10^{-8}\,\mathrm{eV}$.

As well as terrestrial direct searches, a variety of astrophysical observations can be used to constrain the ALP parameter space. For example, Refs.~\cite{Caputo:2018ljp,Caputo:2018vmy,Carenza:2019vzg} examined the possibility of stimulated axion decay for various astrophysical targets, showing that dedicated observations of dwarf spheroidal galaxies can potentially improve upon current constraints by an $\mathcal{O}(1-10)$ factor. Observations of stellar lifetimes also serve as a sensitive probe of ALPs. The hot plasma in the interior of a star readily produces low mass particles which allow for energy transport out of the stellar environment and have the potential to change a stars normal evolution~\cite{Raffelt:1987yu,Raffelt:2006cw,Friedland:2012hj}.

Reference~\cite{Hook:2018iia} recently suggested that the magnetospheres of neutron stars (NSs) can induce enough axion-photon conversion to be subsequently observed by radio telescopes, potentially probing QCD axions.\footnote{This conversion process was originally proposed and studied in Ref.~\cite{Pshirkov:2007st}.} Importantly, the finite electron density of the plasma induces an effective photon mass \cite{Huang:2018lxq} which, when equal to the axion mass, allows for the conversion to become resonant. Since, in the simplest scenarios, the photon mass monotonically decreases with the distance to the NS surface, there exists a continuum of resonant conversion surfaces corresponding to different axion masses. The conversion process also conserves energy (up to Doppler broadening~\cite{Battye:2019aco}), which allows for an inference of the axion mass directly from the radio signal. This is particularly valuable since most terrestrial experiments must fine tune their experimental setups to gain sensitivity to particular axion masses. The two approaches of \textit{direct} and \textit{indirect} detection are therefore highly complementary.

Radio signals from a collection of NSs, such as the bulge population in the centre of our galaxy \cite{Fermi-LAT:2017yoi}, were investigated in Ref.~\cite{Safdi:2018oeu}. Multi-messenger signals (radio and gravitational waves) from Black Hole - Neutron Star inspirals were studied in Ref.~\cite{Edwards:2019tzf}. A detailed study of the mixing equations used to describe axion-photon conversion is provided in Ref.~\cite{Raffelt:1987im,Battye:2019aco}.

In this work we present a more complete treatment of the radio signal calculation for individual NSs. To this end, we extend the recent analytic treatment presented in Ref.~\cite{Hook:2018iia} by performing a numerical ray-tracing computation of the conversion process. This allows us to fully account for the isotropic phase-space distribution of axions in the vicinity of the NS.  We find significant qualitative and quantitative differences with respect Ref.~\cite{Hook:2018iia}, and provide the necessary code to calculate the flux for various parameter combinations. Our results impact the overall reach of future radio searches for QCD axions and ALPs, as well as the optimization of realistic search strategies.

The paper is organized as follows. In \S~\ref{sec:signal} we describe the signal calculation, paying particular attention to the ray-tracing algorithm. In \S~\ref{sec:results}, we report our results for the NS J0806.4-412 and calculate the sensitivity of next-generation radio telescopes to the radio signal and its potential time variability. Finally, we conclude in \S~\ref{sec:concl}.

\section{Signal Calculation \label{sec:signal}}

Here we present the formalism and assumptions behind the ray-tracing method. In particular, we discuss the axion-photon conversion probability, the dark matter distribution, the neutron star's magnetosphere, and the photon flux seen on Earth. 
Where appropriate, we follow the calculations from Ref.~\cite{Hook:2018iia}.

\subsection{Conversion Probability}
 Photons in a plasma acquire an effective mass (``plasma mass") through interactions with free charges, which is given by~\cite{Pshirkov:2007st}
\begin{equation}
    \omega_p = \sqrt{\frac{4\pi \alpha\,n_c}{m_c}}\,,
    \label{eq:plasma}
\end{equation}
where $n_c$ is the charge carrier number density, $m_c$ the charge carrier particle mass, and $\alpha$ is the fine-structure constant. Resonant conversion can occur when the photon plasma mass approximately matches the axion mass $m_a \simeq \omega_p$. This condition singles out a resonant conversion shell which is dependent on the axion mass. Using the WKB and stationary phase approximations, one can show that the probability of an axion converting into a photon while traversing a resonant conversion region is given by
\begin{equation}
  P_{a\to\gamma}=
  \frac{\pi}{2} (g_{a\gamma\gamma} B_{\perp})^2 \frac{1}{v_c |\omega_{p}'|}\,.
  \label{eq:conv_prob}
\end{equation}
Here, $g_{a\gamma\gamma}$ denotes the axion-photon-photon coupling mentioned above and $B_{\perp}$ is the strength of the NS magnetic field perpendicular to the axion trajectory.  The plasma mass derivative $\omega_{p}'$ denotes the derivative along the axion trajectory at the point of resonant conversion.  For the special case of radial trajectories, it is given by $|\omega_p'| = |d\omega_p /dr| = \frac32  m_a/r$ (as in~\cite{Hook:2018iia}).  Further details about the calculation of the conversion probability can be found in Appendix~\ref{apx:prob}, including a critical comparison with earlier literature.

\subsection{Phase Space Distribution of Dark Matter at the Neutron Star's Surface}

The phase-space distribution (PSD), $f(\mathbf{r},\mathbf{v})$, describes the statistical properties (spatial positions, $\mathbf{r}$, and velocities, $\mathbf{v}$) of a group of particles. Assuming that the PSD is stationary gives Liouville's theorem \cite{Liouville:1838zza} from which we can see that the PSD is conserved along the trajectories of the system. We can therefore equate the PSD at infinity to the PSD at the NS surface,
\begin{equation} 
    f(\mathbf{r},\mathbf{v}) = f_{\infty}(\mathbf{r}_{\infty},\mathbf{v}_{\infty})\,,
    \label{eq:Liouvill0e}
\end{equation}
where the subscript infinity refers to quantities far from the NS. The distribution of DM is assumed to be isotropic in the rest frame of the galaxy. The Standard Halo Model predicts an isotropic Maxwellian distribution far from the NS \cite{Piffl:2013mla,Smith:2006ym}, given by
\begin{equation}
    f_{\infty}(\mathbf{r}_{\infty},\mathbf{v}_{\infty}) = \frac{\varrho^{\mathrm{DM}}_{\infty}}{(\pi v_0^2)^{3/2}} \exp\left(-\frac{\mathbf{v}_{\infty}^2}{v_0^2}\right)\,,
    \label{eq:maxwellian}
\end{equation}
where $\varrho^{\mathrm{DM}}_\infty$ is the DM density and $v_0$ is the spread of the distribution (discussed below). Equations~\eqref{eq:Liouvill0e} and \eqref{eq:maxwellian} show that the PSD close to the NS surface will be isotropic as long as $\mathbf{v}_{\infty}^2$ does not depend on the direction of $\mathbf{v}$. We can see this independence directly from energy conservation which allows us to relate the velocity at infinity to the physical velocity seen by a local observer at radius $r$ (the radial coordinate of the Schwarzschild metric) from the NS centre (in the limit $\mathbf{v}_\infty^2 \ll 1$)\footnote{This follows from gravitational redshift $E_\infty^2 = (1-2GM_\text{NS}/r)E_r^2$, where $E_\infty$ and $E_r$ are the axion energy at infinity and radius $r$, respectively.}
\begin{equation}
    \mathbf{v}_{\infty}^2 \simeq \mathbf{v}(r)^2  - \frac{2GM_{\mathrm{NS}} }{r} \,,
\end{equation}
where $G$ is Newton's constant and $M_{\mathrm{NS}}$ is the mass of the NS. The isotropy of the PSD at the NS surface holds as long as the NS velocity is small with respect to the galactic rest frame.\footnote{Note that this is not necessarily a good assumption since the NS and the DM are moving non-relativistically. We will extend this formalism to account for boosts into the NS's reference frame in future work.}
Equations~(10) and~(11) of Ref.~\cite{2006PhRvD..74h3518A} directly yield the local DM PSD:
\begin{equation}
    f(\mathbf{r},\mathbf{v}) = 
    \frac{\varrho^{\mathrm{DM}}_{\infty}}{(\pi v_0^2)^{3/2}} \exp\left(\frac{ 2 GM_{\mathrm{NS}}}{r \, v_0^2}\right) \exp\left(-\frac{\mathbf{v}^2}{v_0^2}\right)\,.
    \label{eq:PSD_Nssurf}
\end{equation}
Note that due to energy conservation, the minimum velocity at a given radius is $v_{\mathrm{min}}(r) = \sqrt{2G M_{\mathrm{NS}} / r}$.  Integrating Eq.~\eqref{eq:PSD_Nssurf} with respect to $\mathbf{v}$ such that $|\mathbf{v}| \geq \sqrt{2 G M_{\mathrm{NS}} / r}$, we obtain:
\begin{equation}
    \varrho(r) = \frac{2 \varrho^{\mathrm{DM}}_{\infty}}{\sqrt{\pi}} \left( x + e^{x^2} \int_{x}^{\infty} e^{-u^{2}} \mathrm{d} u \right)\,,
    \label{eq:rhoDM}
\end{equation}
where $x = \sqrt{2 GM_{\mathrm{NS}}/r v_0^2}$ and $u= v/ v_0$. The DM velocity dispersion $v_0$ is small enough that, in practice, we can take the limit $x \gg 1$ which allows us to neglect the RHS of Eq.~\eqref{eq:rhoDM} and work with the simpler expression:
\begin{equation}
    \varrho(r) = \frac{2 \varrho^{\mathrm{DM}}_{\infty}}{\sqrt{\pi}} \sqrt{\frac{2 GM_{\mathrm{NS}}}{r}} \frac{1}{v_0} \,.
\end{equation}
Furthermore, conservation of energy for infalling axions further yields the DM velocity \begin{equation}
v(r) \simeq \sqrt{\frac{2 G M_{\mathrm{NS}}}{r}} \,.
\end{equation}

In addition to the overall normalisation of the signal, we must worry about the width of the line. We consider two contributions to this width: firstly from the intrinsic velocity distribution of DM far from the neutron star and secondly from the overall Doppler broadening due to the collective motion of the magnetosphere associated with the NS spin. The former is given by the Maxwell-Boltzmann distribution which leads to $\mathcal{B} \sim\left(v_0 / c\right)^{2} m_{a} /(2 \pi)$ where $\mathcal{B}$ is the bandwidth (discussed below). The latter is described in Ref.~\cite{Battye:2019aco}, where we take a conservative approach by setting the conversion surface to be the maximum conversion radius for any given axion mass.\footnote{We also set $\epsilon=1$ in Eq.~(75) of Ref.~\cite{Battye:2019aco}, its largest possible value. This therefore again represents a conservative assumption.}  It is not clear that our approach is a good description of the \textit{true} line width since each pixel contributes some fraction of the total observed flux. In addition, a full calculation of the width would need to consider reflection and transmission components of the signal separately and further investigate the turbulence in the plasma at the scale of the photon wavelength. We leave this to future work and instead show optimistic and conservative sensitivity curves from the two broadening effects.

\subsection{Neutron Star Magnetosphere}\label{sec:NSmagneto}

\begin{figure}[t!]
\includegraphics[width=1.0\linewidth]{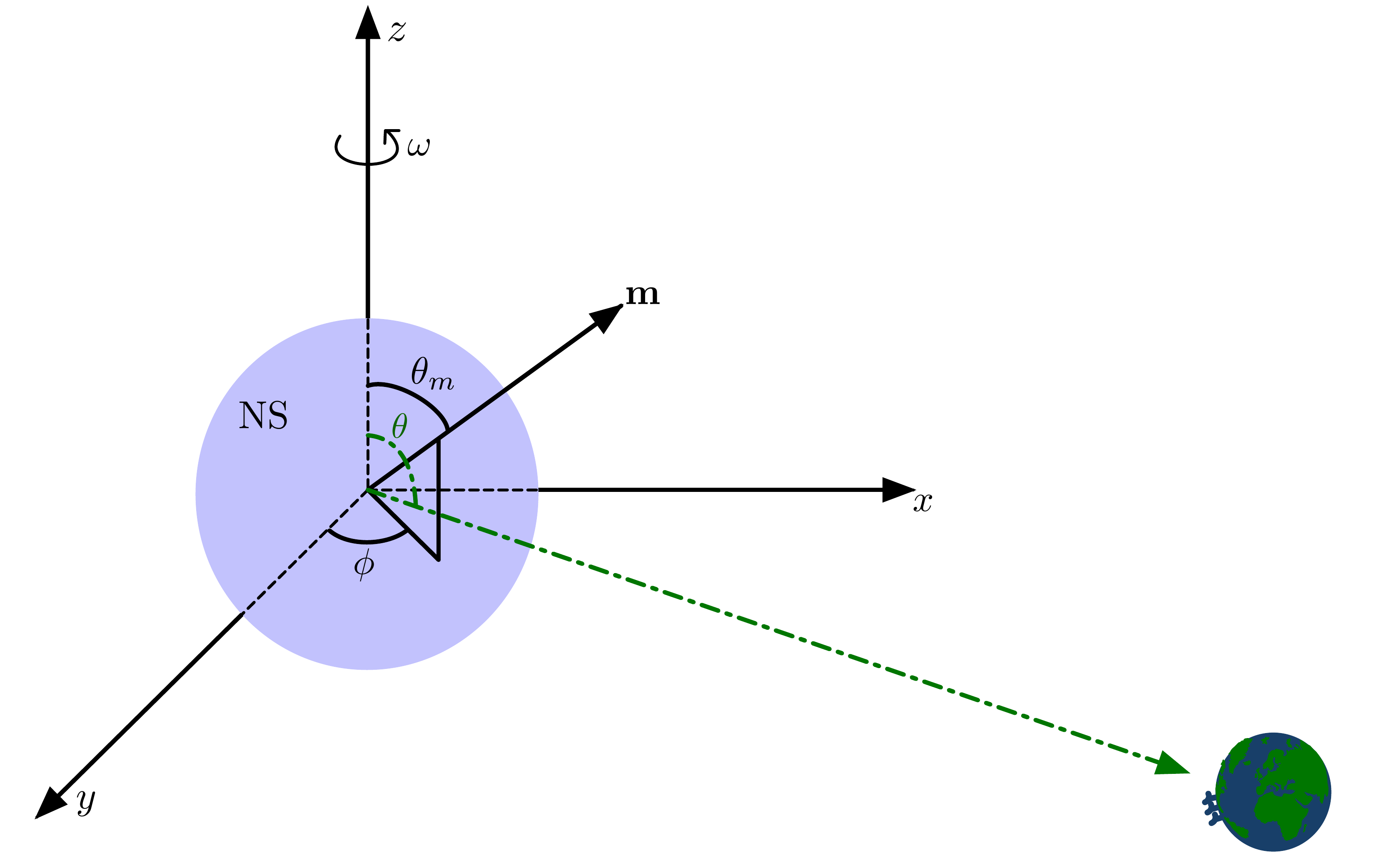}
\caption{{\bf Diagram of the system.}
Here we show a diagram of the NS being observed from Earth with all the relevant quantities labelled for clarity.
\label{fig:angles}}
\end{figure}

As a concrete example, we use the Goldreich and Julian (GJ) model~\cite{1969ApJ...157..869G} as a description of the magnetosphere around the NS (for a review on pulsar magnetosphere see Ref.~\cite{Petri:2016tqe}). The GJ model assumes that the magnetosphere is co-rotating with the NS and provides the following analytic expression for the charge number density:
\begin{equation}
n_{\mathrm{GJ}}(\mathbf{r}) = \frac{2\, \bm{\omega} \cdot \mathbf{B}}{e} \frac{1}{1 - \omega^2 \,r^2\, \sin^2 \theta}\,,
\label{eq:nGJ}
\end{equation}
where $\bm{\omega} = \left(2\pi / P \right) \mathbf{\hat z}$ is the constant NS rotation vector with $P$ being the NS spin period, $\theta$ is the polar angle, and $\mathbf{B}$ is the local magnetic field. The latter is assumed to be in a dipole configuration with axis along the direction $\mathbf{\hat m}$. In spherical coordinates $\left(r,\,\theta,\,\phi\right)$, we have:
\begin{eqnarray}
B_r &=& B_0 \left(\frac{r_{\rm NS}}{r}\right)^3 \left(\cos\theta_m\cos\theta + \sin\theta_m\sin\theta\cos\psi\right) \nonumber\\
B_\theta &=& \frac{B_0}{2} \left(\frac{r_{\rm NS}}{r}\right)^3 \left(\cos\theta_m\cos\theta - \sin\theta_m\cos\theta\cos\psi\right) \\
B_\phi &=& \frac{B_0}{2} \left(\frac{r_{\rm NS}}{r}\right)^3 \sin\theta_m \sin\psi \nonumber \,,
\end{eqnarray}
where $\theta_m$ is the misalignment angle between $\mathbf{\hat z}$ and $\mathbf{\hat m}$, $\psi(t) = \phi - \omega\,t$, $B_0$ is the magnetic field strength at the NS poles, and $r_{\rm NS}$ denotes the NS's radial size. See Fig.~\ref{fig:angles} for a visualisation of the system with the relevant angles labelled.

The plasma mass~\eqref{eq:plasma} is computed by taking $n_c = |n_\mathrm{GJ}|$ and assuming the presence of electrons and positrons only ($m_c = m_e$). By neglecting relativistic corrections (second factor in Eq.~\eqref{eq:nGJ}), we get
\begin{eqnarray}
    \omega_p \left(r,\,\theta,\,\phi;\,t\right) &\simeq& 69.2~{\rm \mu eV} \times \left| 3 \cos\theta \,\mathbf{\hat m}\cdot \mathbf{\hat r} - \cos\theta_m\right|^{1/2} \nonumber\\
    &&\times \left[\frac{B_0}{10^{14}~\mathrm{G}}\, \frac{1~\mathrm{s}}{P} \, \left(\frac{r_\mathrm{NS}}{r}\right)^3\right]^{1/2}\,,
    \label{eq:omega_p_GJ}
\end{eqnarray}
where the time dependence resides in the term
\begin{equation}
    \mathbf{\hat m}\cdot \mathbf{\hat r} = \cos\theta_m\cos\theta + \sin\theta_m\sin\theta\cos\psi(t)\,.
\end{equation}
The resonant axion-photon conversion region is identified by the matching relation $\omega_p \left(r,\,\theta,\,\phi;\,t\right) = m_a$. For a given angle $\theta$, we would expect the radio signal to display some time variation as the NS rotates, due to varying resonant conversion surface observed from Earth. This variation is discussed in Ref.~\cite{Hook:2018iia} where they assumes radial trajectories for the infalling axions. This assumption implies that the radio signal arises from the specific direction $\theta$ connecting the NS to the Earth. For an isotropic PSD, we expect that the absence of any preferred axion trajectory will suppress the time-dependence of the radio signal as it is instead given by the cumulative flux from all possible axion trajectories (as discussed below).

Before concluding, we note that the NS magnetosphere model plays an important role in defining the local properties of the NS plasma. In the present paper, we consider the GJ model with electrons and positrons to make a direct comparison with the analytic results discussed in Ref.~\cite{Hook:2018iia}. Importantly, the GJ model does not describe inhomogeneities in the plasma structure which, in real NSs, may exist. These inhomogeneities may significantly affect the signal and must be accounted for in future work \cite{Carlson:1994yqa}. In addition, the GJ charge density may significantly differ from the true charge density --- this difference is typically described through the multiplicity which can vary significantly \cite{2019ApJ...871...12T}. However, our ray-tracing algorithm is implemented in such a way that the magnetosphere model can be straightforwardly modified to account for more complicated and realistic scenarios capturing local perturbations~\cite{Petri:2016tqe}. The impact of different models (such as the electrosphere model~\cite{10.1093/mnras/213.1.43P} or numerical simulations of the NS magnetosphere~\cite{Philippov:2014mqa,Cerutti:2016ttn,Kalapotharakos:2017bpx}) on the signal prediction is left to future work.

\subsection{Calculating the photon flux}

We assume that axions can convert into photons only at points, denoted $\mathbf{r}_{c}$, where the plasma mass equals the axion mass. The conversion probability $P_{a\to\gamma}(\mathbf{r},\mathbf{u})$ depends on the local magnetic field strength (and therefore the position) as well as the direction of the axion at that point, given by $\mathbf{u}$ with respect to the radial trajectory at $\mathbf{r}_{c}$. Photons entering regions with a larger plasma mass will be reflected, providing a factor of two greater flux when accounting for trajectories parallel and anti-parallel to the line of sight (as mentioned in the next subsection).\footnote{In general, resonant conversion (associated with $k_a^2 = \omega^2-\omega_p^2(r)$) and reflection (associated with $m_a = \omega_p(r)$) do \textit{not} happen at the same place. For radial trajectories, one can show that the difference between both points is given by $\Delta r = \frac{r_c}{3}k_c^2/m_a^2$ (which follows from $\omega_p \propto r^{-3/2}$). The size of the resonant conversion region is given by $L^2 = 2\pi/3\cdot r_c \, k_c / m_a^2$, from which follows that $L \simeq \sqrt{2\pi \Delta r/k_c}$.  We have separation when $\lambda_c \ll \Delta r$, where $\lambda_c$ is the de Broglie wavelength at resonance.} Reflected waves will be will be also Doppler broadened~\cite{Battye:2019aco}, which we will further discuss below. Note that we neglect multiply scattered photons.

The total flux expected from a neutron star can be written in terms of the intensity as
\begin{equation}
F = \int_{\Delta \Omega} \mathrm{d}\Omega\; I(\Omega)\;,
\end{equation}
where the angular integral is over a region \(\Delta \Omega\) which covers the neutron star. We first consider the emission through a planar conversion region (for $a\to\gamma$) that is taken to be perpendicular to the line-of-sight towards the NS, and at a distance \(D\) from the observer.  Its physical size is taken to be infinitesimally small, \(\mathrm{d} A = D^2\, \mathrm{d}\Omega\).  We assume that the conversion region is immersed in an isotropic distribution of axions with number density \(n_a\) and velocity \(v_a\).  The current of axions that traverses the conversion region in direction \(\mathrm{d}\Omega'\) is given by
\begin{equation}
\mathrm{d}^2J(\Omega') = \cos(\alpha)\,\mathrm{d}A\, \frac{\mathrm{d}\Omega'}{4\pi} \,n_a v_a\;,
\label{eqn:dJ}
\end{equation}
where \(\alpha\) is the angle between the line-of-sight from the neutron star to the observer and the direction of the axions. The first two factors give the size of the conversion region projected onto the axion direction. The third and fourth factors together denote the number density of axions moving in direction \(\mathrm{d}\Omega'\), and the last factor is their velocity.  The signal intensity is given by 
\begin{equation}
    I(\Omega) \equiv \frac{\mathrm{d}J^2}{\mathrm{d}\Omega \mathrm{d}A'} =\frac{\mathrm{d}J^2}{\mathrm{d}\Omega' \,\mathrm{d}A} = \frac{n_a v_a}{4\pi}\;.
\end{equation}
As usual \(\mathrm{d}\Omega\) is the angular size of the observed region, and \(\mathrm{d}A'\) is the detector area.  The second equality holds since we can exchange the role of solid angle and observed area (using \(\mathrm{d}\Omega = \mathrm{d}A/D^2\) and \(\mathrm{d}\Omega' = \mathrm{d}A'/D^2\)). In the last step, we set \(\cos(\alpha) \simeq 1\). The \emph{photon} flux from a NS (still assuming a perpendicular conversion region) can therefore be written as 
\begin{equation}
    F = \int_{\Delta \Omega} \mathrm{d}\Omega\frac{n_a v_a}{4\pi}P_{a\to\gamma}
    \simeq \frac{(\Delta b)^2}{D^2} \sum_i \
    \underbrace{\left(\frac{n_a v_a}{4\pi}P_{a\to\gamma}\right)(\Omega_i)}_{\equiv I_\gamma(\Omega_i)}
    \label{eqn:FluxF}
\end{equation}
where we took \(P_{a\to\gamma}\) to be the probability of axion-photon conversion upon traversal of the conversion region, and we split up the integral into sums over different line-of-sights \(i\), which each contribute \(\Delta\Omega_i = \Delta b^2/D^2\) to the integral (and \(\sum_i \Delta\Omega_i = \Delta \Omega\)). Note that these line-of-sights are very close to parallel at the scale of the neutron star.

In order to generalize to non-perpendicular emission planes, the integral over \(\mathrm{d}\Omega\) would have to be, in principle, replaced by an integral over the 2-dim sub-manifold \(\mathcal{C}\). However, we can simply parametrise this sub-manifold by the angular direction seen from the observer, \(\Omega\). This is possible since we only observe (from Earth) the first crossing of the conversion region (everything else is absorbed). In that case, the area of this sub-manifold can be calculated as \(A_{\mathcal{C}} = \int \mathrm{d}A_\mathcal{C} = D^2\int_{\Delta\Omega} \mathrm{d}\Omega \cos(\alpha)^{-1}\;,\) where \(\alpha\) is the angle between the normal of the sub-manifold \(\mathcal{C}\) at each point and the line of sight. The last factor accounts for the deprojection of the sub-manifold when integrating over \(\mathrm{d}\Omega\).  Interestingly, this factor cancels the \(\cos\alpha\) that we obtained in Eq.~\eqref{eqn:dJ}. As a result, the equation for the flux \(F\) in terms of sums over line-of-sights, Eq.~\eqref{eqn:FluxF}, remains the same for non-planar emission regions, provided that the quantities in the parentheses are evaluated at the
point of the conversion.

\subsection{Computational approach}

The total radio flux is computed through a ray-tracing algorithm defined by the following steps:
\begin{enumerate}
    \item We define the region of interest (ROI) as a planar surface $\Delta A$ perpendicular to the line-of-sight towards the NS at a distance $D$ from the Earth and a distance $d$ from the NS centre. The latter is chosen to be the maximum distance at which the resonant axion-photon conversion occurs. Typically, we have $d = \mathcal{O}(100~\mathrm{km})$ for the minimum axion mass considered. The ROI is divided into square pixels of size $\Delta b$, whose centres identify a specific photon trajectory $i$.
    \item For each pixel, we back-propagate the photon by numerically computing its geodesics starting from the centre of the corresponding pixel and taking the initial velocity to be perpendicular to the ROI.
    \item We divide the NS rotation period into intervals of length $\delta t$. For each interval we then compute the plasma mass along each trajectory $i$. The resonant conversion region is determined by numerically solving the equation $\omega_p \left(\bm{r}_{c,i};\,\delta t\right) = m_a$. The position $\bm{r}_{c,i}$ is identified as the first crossing between the photon trajectory and the resonant conversion surface. Other possible crossings correspond to photons with an energy $\omega$ that would have to travel through a plasma with mass $\omega_{p} > \omega$ to reach the observer, and therefore they are scattered during their travel.
    \item The radiated power (which is a useful, distance independent quantity) from each pixel $i$ is then computed at the conversion region as
    \begin{equation}
        \frac{{\rm d}\mathcal{P}_i}{{\rm d} \Omega} = \left\{ \begin{array}{l l}
        2 \times \Delta b^2 \left.\frac{P_{a \rightarrow \gamma}\,m_a n_a\,v_a}{4\pi} \right|_{\bm{r}_{c,i}} &{\rm if}~r_{c,i} \geq r_{\rm NS} \\
        0 &{\rm if}~r_{c,i} < r_{\rm NS}
        \end{array}\right.\,,
    \end{equation}
    where we require that the resonant conversion occurs outside the NS surface. The factor of two takes into account the reflection of photons on the way towards the neutron star. We do not consider the contribution of non-resonant conversion since it is generally sub-dominant.
    \item The total radiated power is obtained by summing the contributions from all the pixels
    \begin{equation}
        \frac{{\rm d}\mathcal{P}}{{\rm d} \Omega} = \sum_i \frac{{\rm d}\mathcal{P}_i}{{\rm d} \Omega} \,,
    \end{equation}
    and the total radio flux is simply given by
    \begin{equation}
        F = \frac{1}{D^2}\frac{{\rm d}\mathcal{P}}{{\rm d} \Omega} \,.
    \end{equation}
    This expression matches the one reported in Eq.~\eqref{eqn:FluxF} multiplied by the axion mass.
\end{enumerate}
Throughout our analysis we consider a flat metric (classical approximation) --- all photons propagate in parallel straight lines. This is a good approximation for the case at hand. The Schwarzschild metric differs from the Minkowski one by terms proportional to $\varepsilon(r) = r_s/r$ with $r_s$ being the Schwarzschild radius. Such a ratio reaches its maximum value at the NS surface where $\varepsilon = \mathcal{O}(0.1)$. We checked that the total radiated power changes by no more than a few percent when considering the Schwarzschild metric.\footnote{Calculating trajectories in the Schwarzchild metric is significantly more computationally expensive than for the flat metric. Since the corrections to the overall signal are small we use the flat metric for efficiency.} There are also additional effects associated with the NS's spin and described by the Kerr metric. This Kerr metric reduces to the Schwarzschild one in the limit of $\varepsilon'(r) = 2 \pi I / P r \ll 1$ where $I$ and $P$ are the moment of inertia and the spin period of the neutron star, respectively. For the isolated NS J0806.4-4123 analyzed in the next section, the maximum value of $\varepsilon'(r)$ at the NS surface is of the order of $10^{-5}$ for reasonable NS moments of inertia~\cite{1994ApJ...424..846R}.

Throughout, we also neglect any general relativistic corrections to our results. Since the Lorentz factors that we encounter for axion velocities can be as large as $\gamma\sim 1.2$, this means that our results are in general only accurate to within $\pm 20\%$ (although the details depend on where most of the observable conversion happens, and we expect much higher accuracy in most cases that are of interest here).

We further checked that the numerical calculation of the radiated power is converged with respect to: {\it i)} the spatial resolution along each trajectory that is used to identify the resonant conversion region; {\it ii)} the pixel resolution of the ROI that sets the number of trajectories. We find that the contribution of each trajectory typically varies within 0.1\% for a spacial resolution of $1~\mathrm{m}$. The pixel resolution is instead fixed by the requirement that the total radio flux does not vary by more than 1\% when increasing the number of trajectories.

\section{Results \label{sec:results}}

Up to now we have not fixed our calculations to a specific system apart from assuming that the magnetosphere is described by the GJ model. The signal is highly dependent on the system being observed. We therefore fix the reference model to be the isolated neutron star J0806.4-4123, which has a period $P=11.37~\mathrm{s}$, a magnetic field $B_{0}=2.5 \times 10^{13}~\mathrm{G}$ ($1~\mathrm{G} = 10^{-4}~\mathrm{T}$), and is located at a distance $D \approx 250~\mathrm{pc}$ from the Earth~\cite{Kaplan:2009ce}. Moreover, we take $M_{\rm NS} = 1 M_\odot$ and $r_{\rm NS} = 10~{\rm km}$ for the mass and the size of the neutron star as well as $\varrho^{\infty}_{\rm DM} = 0.3~{\rm GeV/cm^3}$ and $v_0 = 200~{\rm km / s}$ for the local density and velocity dispersion of DM particles, respectively. In the following, we report the predictions for the radio flux for different angular configurations of the system, and discuss the sensitivity of next-generation radio telescopes to the signal and its time variability. Note that we expect the results of J0806.4-4123 to be qualitatively similar to other NSs. The code we provide is flexible and can be used for any isolated NS system.

\subsection{Flux predictions}
\begin{figure}[t!]
\includegraphics[width=1.0\linewidth]{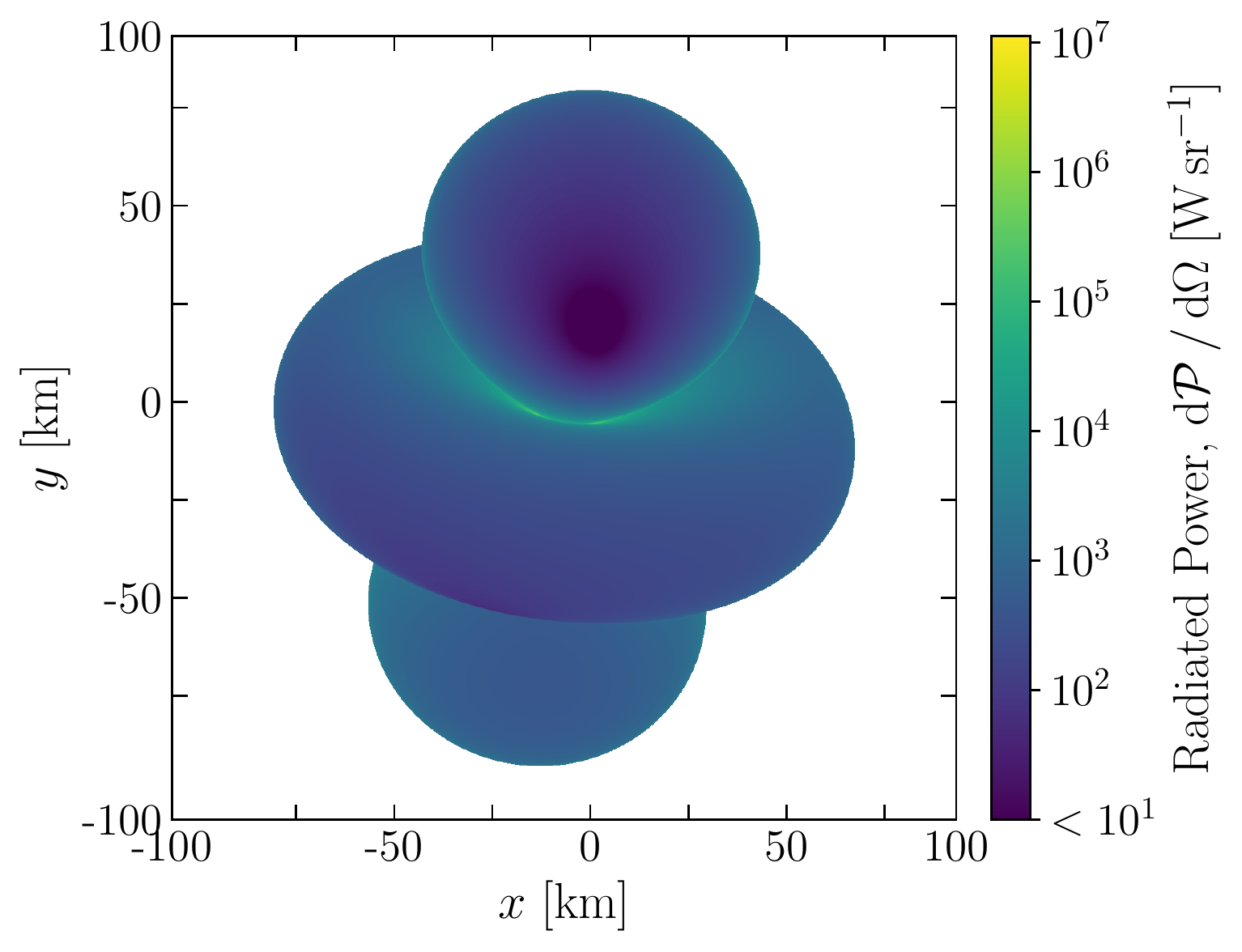}
\caption{{\bf Radiated power from individual trajectories.}
Snapshot of the radiated power from individual trajectories identified by the pixel position $(x,y)$ of the ROI. We consider the benchmark scenario with $\theta_{m} = 15^{\degree}$, $\theta = 58.31^{\degree}$, $g_{a\gamma\gamma} = 1.0 \times 10^{-12}~\mathrm{GeV^{-1}}$, and $m_a = 0.5~\mathrm{\mu eV}$. The video of the evolution of the radiated power during a NS rotation period can be found at: \href{https://youtu.be/VyA1-qbIqB4}{\faFileCodeO}.
\label{fig:rad_pow_trajectories}}
\end{figure}

In Fig.~\ref{fig:rad_pow_trajectories} we show the radiated power of the reference NS from each pixel of the ROI, which corresponds to a planar surface of $100~\mathrm{km} \times 100~\mathrm{km}$ with $6277^2$ pixels ($\Delta b = 30~\mathrm{m}$). We checked that this choice of resolution  indeed meets our convergence criterion. Moreover, we consider the misalignment angle $\theta_{m} = 15^{\degree}$, the direction $\theta = 58.31^{\degree}$ and $\phi=0$. We take the axion-photon coupling $g_{a\gamma\gamma} = 1.0 \times 10^{-12}~\mathrm{GeV^{-1}}$ and the axion mass $m_a = 0.5~\mathrm{\mu eV}$. As can be seen from the plot, there exist very bright pixels corresponding to trajectories for which the conversion region is very close to the NS surface or the crossing with the resonant conversion surface is almost tangential. The former implies that the resonant conversion takes place in regions with very high magnetic fields. The latter, instead, implies that the derivative of the plasma mass along the trajectory is almost zero, therefore strongly enhancing the axion-photon conversion probability. The position of these bright pixels changes during the NS rotation as shown in the video (link is provided in the caption of the figure).
\begin{figure*}[ht!]
    \includegraphics[width=1.0\linewidth]{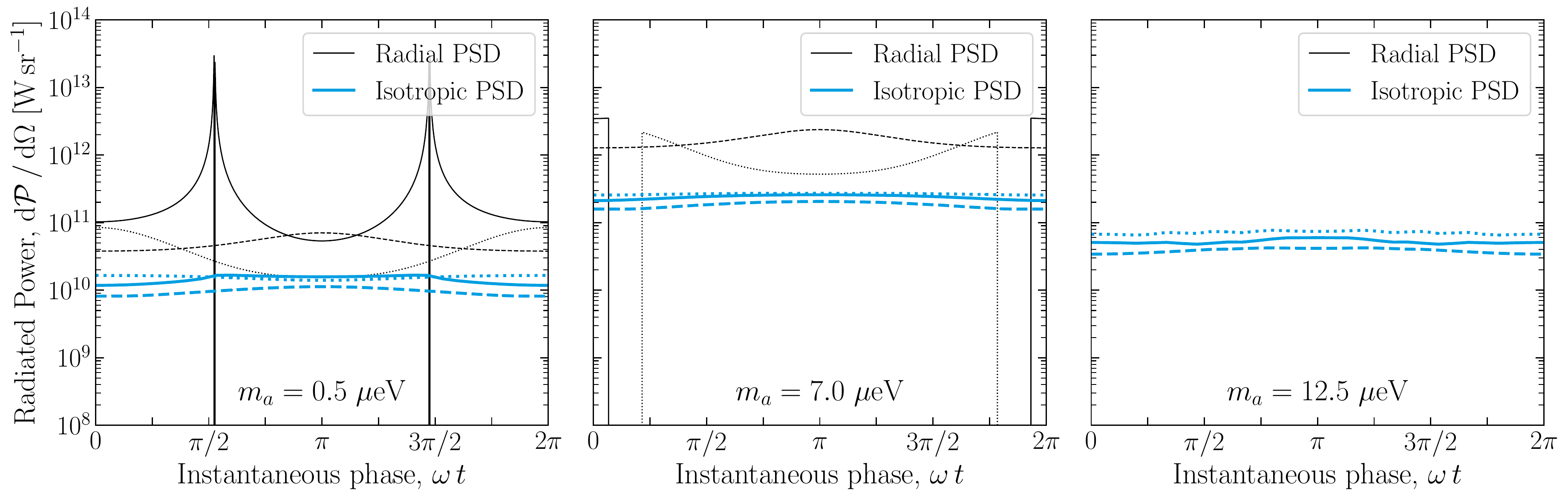}
    \caption{{\bf Total radiated power during a NS rotation period.} Total radiated power as a function of the instantaneous phase ($\omega\,t = 2\pi\,t/P$) from the isolated NS J0806.4-4123, $B_0 = 2.5\times 10^{13}~\mathrm{G}$ and $P = 11.37~\mathrm{s}$. The light blue lines show the ray-tracing results of the present work with an isotropic phase-space distribution of axion particles. The black lines refer to the analytic computation based on radial trajectories~\cite{Hook:2018iia}. The dashed, solid and dotted lines correspond to polar angles $\theta = 36^\circ$, $54^\circ$ and $72^\circ$, respectively. The misalignment angle is fixed to $\theta_m = 18^\circ$ for all lines. The plots from left to right display the radiated power for an axion masses of $0.5~\mathrm{\mu eV}$, $7.0~\mathrm{\mu eV}$ and $12.5~\mathrm{\mu eV}$, respectively.
    \label{fig:rad_pow}}
\end{figure*}

In Fig.~\ref{fig:rad_pow} we report the total radiated power (summing the contribution of all the pixels) as a function of the instantaneous phase during a whole NS period for a few benchmark cases. In particular, the plots from left to right correspond to an increasing axion mass. The solid, dashed, and dotted lines represent the results for a polar angle $\theta$ of $36^\circ$, $54^\circ$ and $72^\circ$, while the azimuthal angle is fixed to $\phi = 0$. Most importantly, our numerical flux predictions (shown in light blue) are compared with the analytic calculations (shown in black) in which the axion PSD is completely radial~\cite{Hook:2018iia}. It is clear that taking into account the contribution of all the trajectories (isotropic PSD) has significant implications. Firstly, we find that an isotropic PSD provides an averaged normalization of the radiated power which is typically up to an order of magnitude smaller than the estimated power from a radial PSD. Secondly, the ray-tracing calculation does not provide a sharp cut-off for the radio signal at large axion masses. The analytic prediction requires the axion-photon conversion to occur outside the NS surface and therefore implies an upper value for the axion mass that can produce a radio signal, as obtained by setting $r_c = r_\mathrm{NS}$ in Eq.~\eqref{eq:omega_p_GJ}. This can be seen by the fact that the black lines are non-zero only in a specific time window during the NS period and, remarkably, are absent in the last plot with $m_a = 12.5~\mathrm{\mu eV}$. Our numerical calculation therefore allows one to extend the radio sensitivity curves to larger axion masses. Thirdly, the isotropic PSD erases the majority of the time variability of the radio signal. The light blue lines are indeed almost flat, while the black curves have large peaks for specific instantaneous phases, especially for low axion masses (see first plot).
\begin{figure*}[ht!]
    \includegraphics[width=1.0\linewidth]{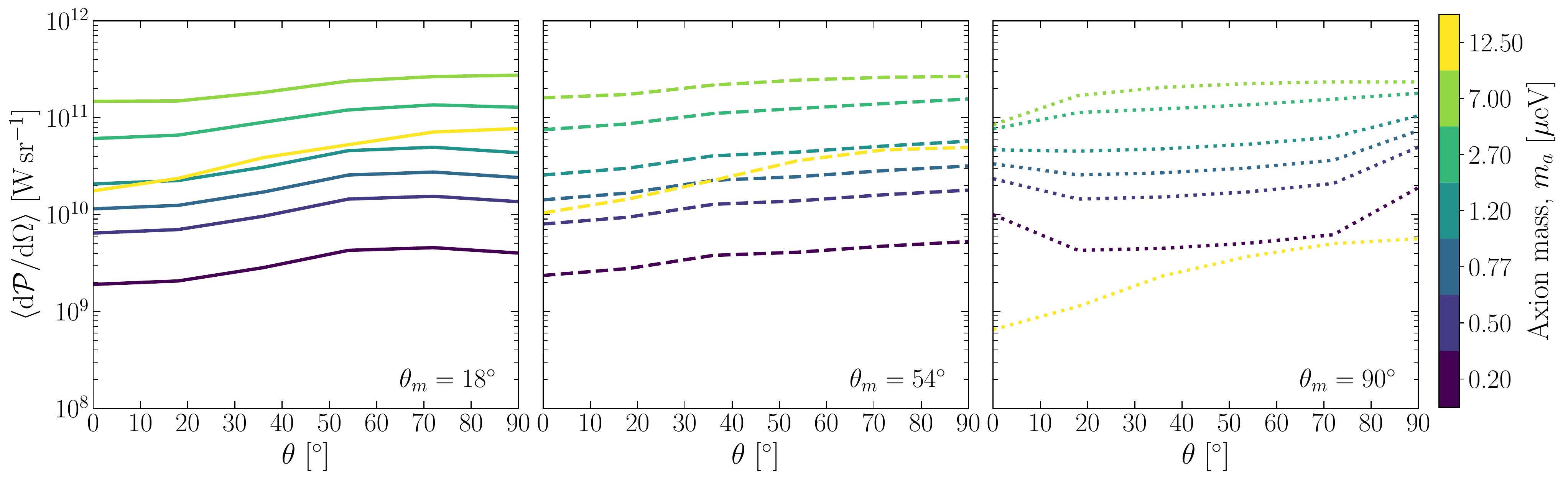}
    \caption{{\bf Averaged radiated power.} Total radiated power averaged over a NS period as a function of the polar viewing angle $\theta$. The three plots from left to right correspond to misalignment angles of $18^\circ$, $54^\circ$ and $90^\circ$, respectively. The colors of the lines refer to different values of the axion mass.
    \label{fig:Mean}}
\end{figure*}
\begin{figure*}[ht!]
    \includegraphics[width=1.0\linewidth]{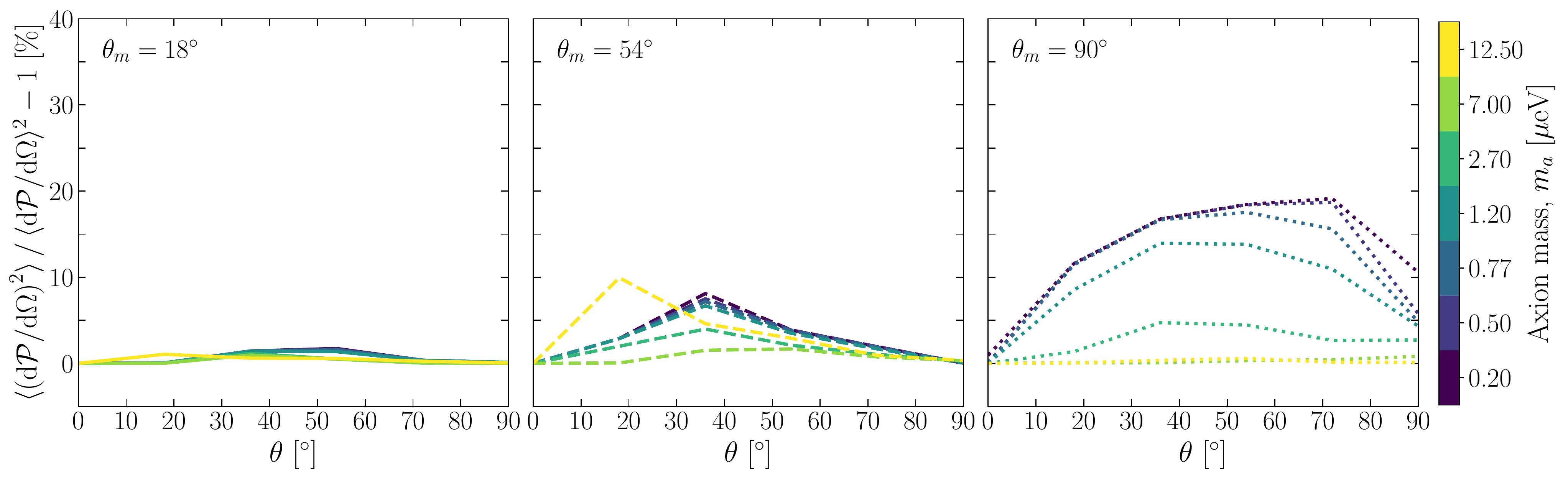}
    \caption{{\bf Relative variance of the radiated power.} Relative variance of the total radiated power with respect to a whole NS period as a function of the polar viewing angle $\theta$. The three plots from left to right correspond to misalignment angles of $18^\circ$, $54^\circ$ and $90^\circ$, respectively. The colors of the lines refer to different values of the axion mass.
    \label{fig:Variance}}
\end{figure*}

Figures~\ref{fig:Mean} and~\ref{fig:Variance} display the average and the relative variance of the radiated power over a NS period as a function of the polar angle $\theta$. The three plots correspond to three different misalignment angles $\theta_m$, while the values of the axion mass are represented by different colors. We note that the averaged radiated power increases as one considers larger axion masses up to a certain value where it starts to decrease, as can been seen for $m_a = 12.5~\mathrm{\mu eV}$ (yellow line). For axion masses larger than a certain threshold, fewer trajectories have a crossing point outside the NS surface and, consequently, cannot contribute to the total radio signal. However, we highlight once again that this is not a sharp change as occurs in the analytic derivation where the radio signal suddenly becomes zero~\cite{Hook:2018iia}. Figure~\ref{fig:Variance} shows that the relative variance of the radio signal significantly depends on the misalignment angle, reaching 30\% for the extreme value $\theta_m = 90^\circ$ (last plot). For reasonably small values of $\theta_m$, the relative variance is instead practically equal to one, implying an almost negligible time variability of the signal.

\subsection{Radio Sensitivity}

Figure~\ref{fig:Variance} shows that the variability of the signal as a function of time is small for realistic values of the misalignment and viewing angles. As discussed above, the time variation of the signal has been reduced (in the majority of cases) to $\mathcal{O}(0-5)\%$. We therefore forecast the sensitivity of radio telescopes to a line detection, although we discuss how well these searches can find the remaining time variability.

\begin{figure}[t!]
\includegraphics[width=1.0\linewidth]{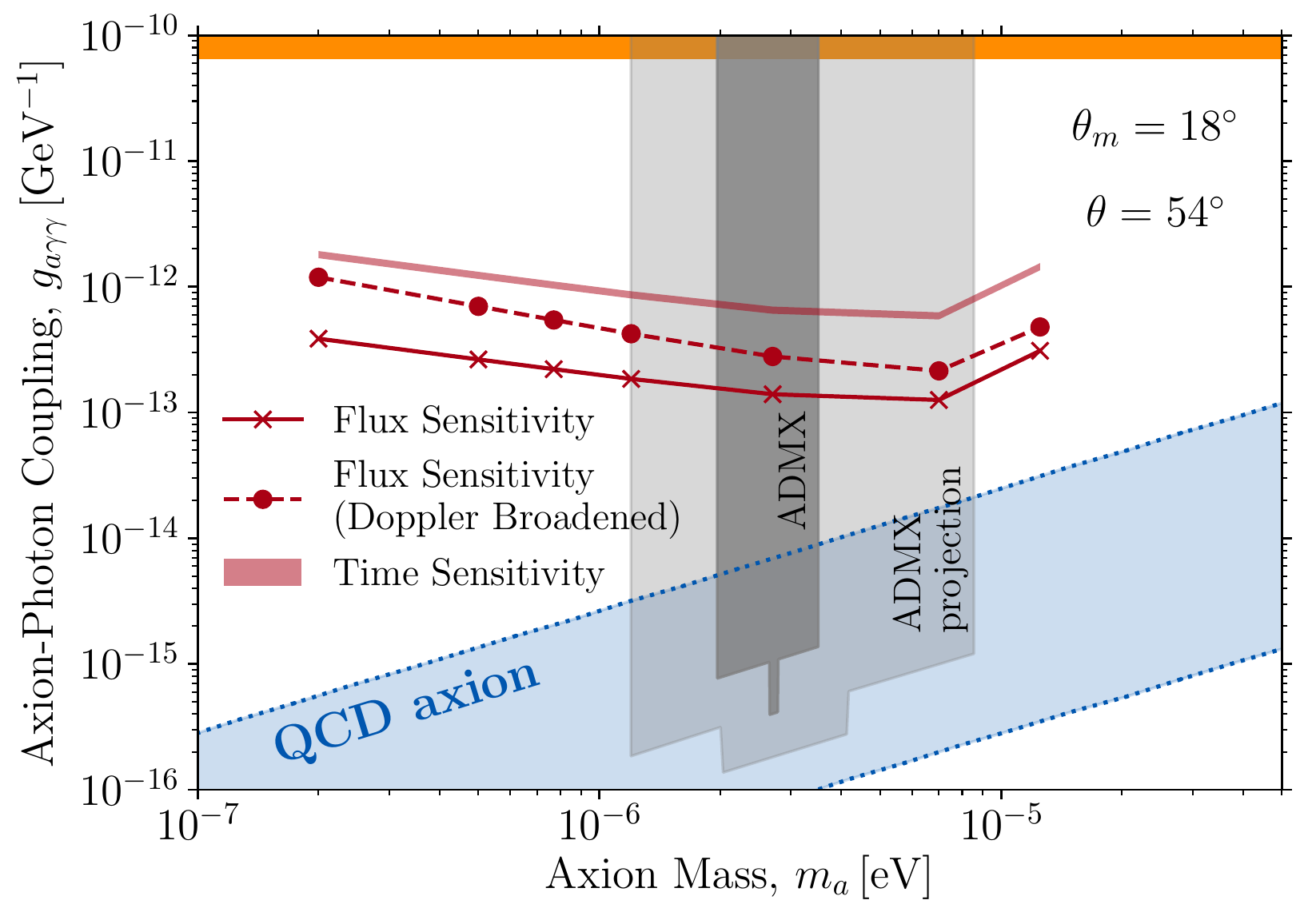}
\caption{{\bf Projected sensitivity to the axion-photon coupling from radio observations.} We consider the isolated NS J0806.4-412 and assume $\tau_{\rm obs} = \mathrm{100\,hrs}$. The two red lines correspond to the sensitivity limit for two line broadening scenarios as described in the text. The red solid line only accounts for the DM velocity distribution far from the NS where as the red dashed line also accounts for Doppler broadening from the rotation of the NS magnetosphere. The red band shows the minimum coupling required to detect the time variation of the signal (here we neglect Doppler broadening).}
\label{fig:sensitivity}
\end{figure}

Assuming the thermal noise of the radio telescope is Gaussian, the signal-to-noise ratio (SNR) is given by
\begin{equation}
\label{eqn:SNR}
    \mathrm{SNR_{L}} = S_{a\rightarrow\gamma} \frac{\sqrt{2\,\mathcal{B}\,\tau_{\rm obs}}}{\rm SEFD}\,,
\end{equation}
where $S_{a\rightarrow\gamma}$ is the flux density of the source, $\rm{SEFD}$ is the system equivalent flux density, $\mathcal{B}$ bandwidth of the signal, $\tau_{\rm obs}$ the observation time, and the factor of two simply accounts for the number of polarizations. The flux density is given by $S_{a\rightarrow\gamma} = F/\mathcal{B}$. The bandwidth is described in \S~\ref{sec:signal} where we consider two contributions to the broadening of the line. The sensitivity to both scenarios is shown in Fig.~\ref{fig:sensitivity} where the solid red line corresponds to broadening from the DM velocity distribution only and the red dashed line accounts for Doppler broadening from the rotation of the magnetosphere as well.

Realistically, the radio sensitivity will be limited by a telescopes ability to remove false positives. For example, interference from radio sources on Earth or satellites in the field of view can produce radio line emission. A confirmation of the signal therefore requires an interferometer capable of taking ON/OFF spectra of the source in radio quiet regions. We therefore consider the future telescope SKA which is projected to have $\rm SEFD \sim0.098~{\rm Jy}$~\cite{SKA} and assume an observation time of $\tau_{\rm obs} = \mathrm{100\,hrs}$. Figure~\ref{fig:sensitivity} shows the $2\,\sigma$ sensitivity\footnote{Note that Ref.~\cite{Hook:2018iia} uses $1\,\sigma$ sensitivity which is too low for discovery.} to the radio line for various masses. In addition, we show the region of parameter space that could correspond to the QCD axion (blue), the current and future ADMX sensitivity in dark and light grey respectively~\cite{Asztalos:2009yp,Du:2018uak}, and the CAST sensitivity in orange~\cite{Anastassopoulos:2017ftl}. 

Although reduced, the time variability of the radio line would provide a striking confirmation of its astrophysical origin. We therefore estimate the scaling from a $2\, \sigma$ measurement of the line to a $5\,\sigma$ detection of the variability. We therefore make the substitution
\begin{equation}
S_{a\rightarrow\gamma} \rightarrow \sqrt{\langle S_{a\rightarrow\gamma}^2\rangle - \langle S_{a\rightarrow\gamma}\rangle^2}\sim\sqrt{\sigma_\mathrm{T}}\,S_{a\rightarrow\gamma}\,,
\end{equation}
where $\sigma_\mathrm{T}$ denotes the variance of the signal over a NS rotation. From Fig.~\ref{fig:Variance} we see that, for $\theta_m = 18^{\circ}$, we have $0.0102\lesssim \sigma_\mathrm{T}\lesssim 0.0172$ depending on the axion mass. We can then scale line detection sensitivity to the smallest coupling with a detectable time variability as
\begin{equation}
    g_{a\gamma\gamma}^\mathrm{T} = \left(\frac{\mathrm{SNR_{T}}}{\mathrm{SNR_{L}}\sqrt{\sigma_\mathrm{T}}}\right)^{1/2} g_{a\gamma\gamma}^\mathrm{L} \, ,
\end{equation}
where $\mathrm{SNR_{T}}$ is the signal-to-noise ratio for a detection of the time variation which we set to 5. The red band in Fig.~\ref{fig:sensitivity} shows this scaling for the range of values for $\sigma_\mathrm{T}$. In particular, we can see that a significantly larger coupling (and subsequently larger flux) is required to detect the small variation in the signal. For larger values of $\theta_m$ we see greater variation in the signal, making it easier to detect, although this effect is still greatly suppressed with respect to Ref.~\cite{Hook:2018iia}.

\section{Discussion and Conclusions\label{sec:concl}}

In this work we have calculated the expected flux from axion-photon conversion in a NS magnetosphere. To correctly account for the isotropic phase space distribution of the DM at the NS surface we performed a ray-tracing procedure, back propagating photons to the conversion surface to find the total flux at Earth. Our work builds upon Ref.~\cite{Hook:2018iia} with two primary results:
\begin{itemize}
    \itemsep0em 
    \item The predicted flux is large enough to potentially detects ALPs down to $g_{a\gamma\gamma}\sim 10^{-12} \,\mathrm{GeV}^{-1}$;
    \item The time variation of the signal is small for realistic values of the misalignment angle, $0^\circ\lesssim \theta_m \lesssim 30^\circ$, where we see, at maximum, an $\mathcal{O}(5\%)$ modulation of the signal.
\end{itemize}
Both of these effects can be seen in Fig.~\ref{fig:rad_pow} in which the blue lines indicate this work and the black lines show the radial approximation of Ref.~\cite{Hook:2018iia} where the time variation is greatly enhanced. We note however that our analysis currently neglects two important physical effects: reflection and refraction of photons escaping after the conversion process. Both of these effects could change the magnitude and time variability of the final signal. Nevertheless, we have performed the most detailed calculation of the signal to date and show that the flux is still observable for a variety of angular configurations, as shown in Fig.~\ref{fig:Mean}. Figure~\ref{fig:Variance} shows the variance of the signal for different viewing and misalignment angles from which it is clear that only for unrealistically large misalignment angles, $\theta_m\gtrsim30^\circ$, do we see an appreciable modulation of the signal. 

In Fig.~\ref{fig:sensitivity} we show that observations of NS targets, such as J0806.4-4123, with future telescopes such as SKA will probe unexplored regions of the ALP parameter space. Although the time variability of the signal is small compared to previous estimates, a measurement of this variability would provide a striking signature of the signal's astrophysical origin. We therefore estimate the coupling required to make a $5\,\sigma$ detection of the time variation for $\theta_m=18^\circ$ and $\theta=54^\circ$, showing that observations J0806.4-4123 could still detect this variation for couplings below the CAST limit.

By accounting for the isotropic phase space distribution of DM we are able to extend the sensitivity of SKA to higher axion masses than Ref.~\cite{Hook:2018iia}. The high mass cut off of the sensitivity is set by the requirement that the conversion process occurs outside the NS interior which, in our setup, occurs at a different mass for each pixel. At high axion masses we therefore retain a fraction of the overall flux induced by non-radial trajectories. This effect is also reflected by the reduction of sensitivity at high masses $m_a\gtrsim 10^{-5}\, \mathrm{eV}$, as seen in Fig.~\ref{fig:sensitivity}.

Although we have made a crucial step towards calculating the true signal, there are a number of caveats that should be addressed in future work. 
Firstly, we neglect the boost of the NS with respect to the galactic rest frame. In practice this boost would mean that the NS sees a prevailing DM \textit{wind}, similar to the wind studied in direct detection experiments \cite{PhysRevD.33.3495}. Accounting for this effect is relatively simple if the boost is known but we leave this to future work.
Secondly, we assume that the GJ model is a good approximation of the NS magnetosphere which, for realistic NSs, may not be the case. Future work should systematically understand how realistic NS magnetosphere models can effect both the magnitude and width of the radio line. Although this may only be possible with full 3D simulations, a systematic study of different analytic models would provide valuable information for the space of possible signatures. As mentioned in Sec.~\ref{sec:NSmagneto}, inhomogeneities in the plasma can significantly affect the signal calculation and should also be accounted for in future work. In addition, the multiplicity of the charge density can vary significantly from the GJ value, potentially changing the conversion region's size and its distance from the NS. Reference~\cite{Battye:2019aco} recently studied the Doppler broadening of the line due to the motion of the magnetosphere, showing that there is potentially a significant contribution to the overall width. Moving forward, it is important that the precise width and its evolution in time is computed more accurately, accounting for turbulence in the plasma as well as the transmission and reflection components of the signal. In particular, multiply scattered photons have to be taken into account consistently. This is especially important for conversion in the \textit{throat} of the NS magnetosphere as the photon production can be greatly enhanced in this region but may not escape to infinity without scattering. We leave this to future work. Overall, we have taken a step towards understanding the true signal of axion-photon conversion from a NS --- future work will build upon our framework by incorporating many of the physical effects mentioned above and reducing the number of assumptions we made in this work.

Finally, we emphasise the complementarity between indirect and direct searches for axion DM. Given a detection of a radio line from a NS it would be easy to confirm its DM nature through measurements with cavity searches such as ADMX. Importantly, the frequency of the radio signature would allow for an inference of the axion mass and subsequently reduce the frequency range through which direct experiments would need to search (a primary issue in direct searches for axions in resonant cavities). Both \textit{direct} and \textit{indirect} approaches therefore represent fundamental tools in the search for axion dark matter. Code used for the calculations throughout this work can be found at \href{https://github.com/mikaelLEROY/AxionNS_RayTracing}{\faGithub}.

\medskip
\acknowledgments{
We thank the python scientific computing packages numpy~\cite{numpy} and scipy~\cite{scipy}. This research is funded by NWO through the VIDI research program ``Probing the Genesis of Dark Matter" (680-47-532; TE, CW). TE acknowledges support by the Vetenskapsr\r{a}det (Swedish Research Council) through contract No. 638-2013-8993.
}

\bibliography{main}

\onecolumngrid

\appendix 

\input{appendix}

\end{document}

%% file: appendix.tex
\section{Derivation of the Axion-Photon Conversion Probability}
\label{apx:prob}

Here we briefly sketch our derivation of the conversion probability. Related discussions can be found in~\cite{Hook:2018iia} and~\cite{Battye:2019aco} (see also Ref.~\cite{Raffelt:1987im}). Since the conversion probability derived in these references differs by a factor of $1/v_c$ (with $v_c$ being the axion velocity at the conversion point), here we reproduce our reasoning for adopting the results from~\cite{Battye:2019aco} for our signal calculations.

The Lagrangian for a system of axions $a$ and photons $A^\mu$ reads
\begin{equation}
\label{eqn:lag}
\mathcal{L}(A_{\nu},\phi) = - \frac{1}{4} F^{\mu \nu} F_{\mu \nu}  - A_{\nu} J^{\nu} -\frac{1}{2} m_{a}^2 a^2 + \frac{1}{2} \partial_{\mu} a \partial^{\mu}a - \frac{1}{4} g_{a \gamma \gamma} F_{\mu\nu} \tilde{F}^{\mu\nu} a\;,
\end{equation}
where $m_a$ is the axion mass, and $g_{a\gamma\gamma}$ the axion-photon-photon coupling.  

Photons propagating in a plasma acquire an effective mass $\omega_{p}(\mathbf{r})$ due to their interactions with free charges. 
Considering axion-photon fields propagating only along the z direction in an intense transverse magnetic field the relevant equations of motion that can be derived from Eq.~\eqref{eqn:lag} via the Euler-Lagrange equations are
\begin{align}
(\partial_{t}^2 - \partial_{z}^2 + \omega_{p}^{2}) \, A_{x}(z,t) &= - \beta(z) \partial_{t} a(z,t),
\\
(\partial_{t}^2 - \partial_{z}^2 + m_{a}^2) \, a(z,t) &= - \beta(z) \partial_{t} A_{x}(z,t)\;,
\end{align}
where we defined $\beta(z) \equiv g_{a\gamma\gamma} B_{x}(z)$.

For a plane wave scalar field $a$, the energy flux (derived from the time/space component of the stress-energy tensor, and conserved in the absence of conversion) is given by $f_a \propto k_a |a|^2$, where $k_a$ is the momentum of the field, and an equivalent expression holds for the photon field. The probability of an axion to convert into a photon after a distance $z$ can therefore be written as 
\begin{equation}
    P_{a\to\gamma} := \frac{k_\gamma(z)^2}{k_a(0)^2}\left| \frac{A_{x}(z,t)}{a(0,t)} \right|^2\;,
\label{eq:conversionProbability}
\end{equation}
where $k_\gamma$ is the photon momentum.
Note that we only consider the transmitted wave and ignore reflections for now.
Using a WKB approximation, one can show that the photon field amplitude takes the form (up to factors of $e^{-i\omega t}$)
\begin{equation}
    A_{x}(z) = -\frac{\omega}{2}a(0)\;e^{i \int_{0}^{z} k_{\gamma}(z') \mathrm{d} z'} \int_{0}^{z} \frac{ \beta(z')}{\sqrt{k_\gamma(z)k_\gamma(z')}} e^{  i \int_{0}^{z'} (k_a - k_{\gamma}(z'')) \mathrm{d} z''} \mathrm{d} z'\;.
\end{equation}
We are interested in the resonant conversion of axions (when $\omega_{p} \simeq m_{a}$) since the conversion probability is maximised here. We therefore use the stationary phase approximation to evaluate the integrals.  Expanding the photon mass $\omega_{p}$ around the axion mass $m_a$ to first order, the argument of the exponential on the right-hand side becomes
$k_a - k_{\gamma}(z) \simeq  - k_\gamma'(z_0) \eta \simeq -m_{a} \, \eta \, \omega_{p}(z_0)' / \sqrt{\omega^2 - m_{a}^2}$, where $\eta \equiv z - z_{0}$ (resonant conversion occurs at $z_0$). Assuming a constant magnetic field throughout the conversion region, the resonant forward conversion probability for axions into photons can then be approximated as
\begin{equation}
    P_{a\to\gamma} \simeq 
    \left(\frac{g \omega}{2 k_\gamma}\right)^2
    \left| \int_{-\eta}^{\eta} \mathrm{d}\eta' B(\eta') e^{-i \int_{-\eta}^{\eta'} \mathrm{d}\eta'' k_\gamma'(z_0) \eta''}\right|^2 \;.
\end{equation}
If we assume that the magnetic field varies only slowly around the conversion region, the integrals can be evaluated analytically and one obtains (in the limit $\eta \to \infty$)
\begin{equation}
    \label{eqn:final_conversion}
    P_{a\to\gamma} = 
    \frac{\pi}{2|k_\gamma'|}  \left( \frac{g B \omega}{k_a} \right)^2 \simeq
    \frac{\pi (g B)^2}{2|\omega_p'|v_a} \;,
\end{equation}
where all quantities are evaluated at the conversion point. Equation~\eqref{eqn:final_conversion} is equivalent to the expressions obtained in~\cite{Battye:2019aco}, and differs from Ref.~\cite{Hook:2018iia} by an additional factor $1/v_a$ (which enhances to overall emission). Note that throughout we neglect corrections of the order of the Lorentz factor and always approximate $\gamma \simeq 1$ (which is accurate to within $\lesssim 20\%$ in our scenarios).

%% file: main.bbl
\begin{thebibliography}{73}%
\makeatletter
\providecommand \@ifxundefined [1]{%
 \@ifx{#1\undefined}
}%
\providecommand \@ifnum [1]{%
 \ifnum #1\expandafter \@firstoftwo
 \else \expandafter \@secondoftwo
 \fi
}%
\providecommand \@ifx [1]{%
 \ifx #1\expandafter \@firstoftwo
 \else \expandafter \@secondoftwo
 \fi
}%
\providecommand \natexlab [1]{#1}%
\providecommand \enquote  [1]{``#1''}%
\providecommand \bibnamefont  [1]{#1}%
\providecommand \bibfnamefont [1]{#1}%
\providecommand \citenamefont [1]{#1}%
\providecommand \href@noop [0]{\@secondoftwo}%
\providecommand \href [0]{\begingroup \@sanitize@url \@href}%
\providecommand \@href[1]{\@@startlink{#1}\@@href}%
\providecommand \@@href[1]{\endgroup#1\@@endlink}%
\providecommand \@sanitize@url [0]{\catcode `\\12\catcode `\$12\catcode
  `\&12\catcode `\#12\catcode `\^12\catcode `\_12\catcode `\%12\relax}%
\providecommand \@@startlink[1]{}%
\providecommand \@@endlink[0]{}%
\providecommand \url  [0]{\begingroup\@sanitize@url \@url }%
\providecommand \@url [1]{\endgroup\@href {#1}{\urlprefix }}%
\providecommand \urlprefix  [0]{URL }%
\providecommand \Eprint [0]{\href }%
\providecommand \doibase [0]{http://dx.doi.org/}%
\providecommand \selectlanguage [0]{\@gobble}%
\providecommand \bibinfo  [0]{\@secondoftwo}%
\providecommand \bibfield  [0]{\@secondoftwo}%
\providecommand \translation [1]{[#1]}%
\providecommand \BibitemOpen [0]{}%
\providecommand \bibitemStop [0]{}%
\providecommand \bibitemNoStop [0]{.\EOS\space}%
\providecommand \EOS [0]{\spacefactor3000\relax}%
\providecommand \BibitemShut  [1]{\csname bibitem#1\endcsname}%
\let\auto@bib@innerbib\@empty
\bibitem [{\citenamefont {Peccei}\ and\ \citenamefont
  {Quinn}(1977{\natexlab{a}})}]{Peccei:1977hh}%
  \BibitemOpen
  \bibfield  {author} {\bibinfo {author} {\bibfnamefont {R.~D.}\ \bibnamefont
  {Peccei}}\ and\ \bibinfo {author} {\bibfnamefont {H.~R.}\ \bibnamefont
  {Quinn}},\ }\href {\doibase 10.1103/PhysRevLett.38.1440} {\bibfield
  {journal} {\bibinfo  {journal} {Phys. Rev. Lett.}\ }\textbf {\bibinfo
  {volume} {38}},\ \bibinfo {pages} {1440} (\bibinfo {year}
  {1977}{\natexlab{a}})},\ \bibinfo {note} {[,328(1977)]}\BibitemShut {NoStop}%
\bibitem [{\citenamefont {Peccei}\ and\ \citenamefont
  {Quinn}(1977{\natexlab{b}})}]{Peccei:1977ur}%
  \BibitemOpen
  \bibfield  {author} {\bibinfo {author} {\bibfnamefont {R.~D.}\ \bibnamefont
  {Peccei}}\ and\ \bibinfo {author} {\bibfnamefont {H.~R.}\ \bibnamefont
  {Quinn}},\ }\href {\doibase 10.1103/PhysRevD.16.1791} {\bibfield  {journal}
  {\bibinfo  {journal} {Phys. Rev.}\ }\textbf {\bibinfo {volume} {D16}},\
  \bibinfo {pages} {1791} (\bibinfo {year} {1977}{\natexlab{b}})}\BibitemShut
  {NoStop}%
\bibitem [{\citenamefont {Weinberg}(1978)}]{Weinberg:1977ma}%
  \BibitemOpen
  \bibfield  {author} {\bibinfo {author} {\bibfnamefont {S.}~\bibnamefont
  {Weinberg}},\ }\href {\doibase 10.1103/PhysRevLett.40.223} {\bibfield
  {journal} {\bibinfo  {journal} {Phys. Rev. Lett.}\ }\textbf {\bibinfo
  {volume} {40}},\ \bibinfo {pages} {223} (\bibinfo {year} {1978})}\BibitemShut
  {NoStop}%
\bibitem [{\citenamefont {Wilczek}(1978)}]{Wilczek:1977pj}%
  \BibitemOpen
  \bibfield  {author} {\bibinfo {author} {\bibfnamefont {F.}~\bibnamefont
  {Wilczek}},\ }\href {\doibase 10.1103/PhysRevLett.40.279} {\bibfield
  {journal} {\bibinfo  {journal} {Phys. Rev. Lett.}\ }\textbf {\bibinfo
  {volume} {40}},\ \bibinfo {pages} {279} (\bibinfo {year} {1978})}\BibitemShut
  {NoStop}%
\bibitem [{\citenamefont {Preskill}\ \emph {et~al.}(1983)\citenamefont
  {Preskill}, \citenamefont {Wise},\ and\ \citenamefont
  {Wilczek}}]{Preskill:1982cy}%
  \BibitemOpen
  \bibfield  {author} {\bibinfo {author} {\bibfnamefont {J.}~\bibnamefont
  {Preskill}}, \bibinfo {author} {\bibfnamefont {M.~B.}\ \bibnamefont {Wise}},
  \ and\ \bibinfo {author} {\bibfnamefont {F.}~\bibnamefont {Wilczek}},\ }\href
  {\doibase 10.1016/0370-2693(83)90637-8} {\bibfield  {journal} {\bibinfo
  {journal} {Phys. Lett.}\ }\textbf {\bibinfo {volume} {120B}},\ \bibinfo
  {pages} {127} (\bibinfo {year} {1983})}\BibitemShut {NoStop}%
\bibitem [{\citenamefont {Abbott}\ and\ \citenamefont
  {Sikivie}(1983)}]{Abbott:1982af}%
  \BibitemOpen
  \bibfield  {author} {\bibinfo {author} {\bibfnamefont {L.~F.}\ \bibnamefont
  {Abbott}}\ and\ \bibinfo {author} {\bibfnamefont {P.}~\bibnamefont
  {Sikivie}},\ }\href {\doibase 10.1016/0370-2693(83)90638-X} {\bibfield
  {journal} {\bibinfo  {journal} {Phys. Lett.}\ }\textbf {\bibinfo {volume}
  {120B}},\ \bibinfo {pages} {133} (\bibinfo {year} {1983})}\BibitemShut
  {NoStop}%
\bibitem [{\citenamefont {Dine}\ and\ \citenamefont
  {Fischler}(1983)}]{Dine:1982ah}%
  \BibitemOpen
  \bibfield  {author} {\bibinfo {author} {\bibfnamefont {M.}~\bibnamefont
  {Dine}}\ and\ \bibinfo {author} {\bibfnamefont {W.}~\bibnamefont
  {Fischler}},\ }\href {\doibase 10.1016/0370-2693(83)90639-1} {\bibfield
  {journal} {\bibinfo  {journal} {Phys. Lett.}\ }\textbf {\bibinfo {volume}
  {120B}},\ \bibinfo {pages} {137} (\bibinfo {year} {1983})}\BibitemShut
  {NoStop}%
\bibitem [{\citenamefont {Klaer}\ and\ \citenamefont
  {Moore}(2017)}]{Klaer:2017ond}%
  \BibitemOpen
  \bibfield  {author} {\bibinfo {author} {\bibfnamefont {V.~B.}\ \bibnamefont
  {Klaer}}\ and\ \bibinfo {author} {\bibfnamefont {G.~D.}\ \bibnamefont
  {Moore}},\ }\href {\doibase 10.1088/1475-7516/2017/11/049} {\bibfield
  {journal} {\bibinfo  {journal} {JCAP}\ }\textbf {\bibinfo {volume} {1711}},\
  \bibinfo {pages} {049} (\bibinfo {year} {2017})},\ \Eprint
  {http://arxiv.org/abs/1708.07521} {arXiv:1708.07521 [hep-ph]} \BibitemShut
  {NoStop}%
\bibitem [{\citenamefont {Gorghetto}\ \emph {et~al.}(2018)\citenamefont
  {Gorghetto}, \citenamefont {Hardy},\ and\ \citenamefont
  {Villadoro}}]{Gorghetto:2018myk}%
  \BibitemOpen
  \bibfield  {author} {\bibinfo {author} {\bibfnamefont {M.}~\bibnamefont
  {Gorghetto}}, \bibinfo {author} {\bibfnamefont {E.}~\bibnamefont {Hardy}}, \
  and\ \bibinfo {author} {\bibfnamefont {G.}~\bibnamefont {Villadoro}},\ }\href
  {\doibase 10.1007/JHEP07(2018)151} {\bibfield  {journal} {\bibinfo  {journal}
  {JHEP}\ }\textbf {\bibinfo {volume} {07}},\ \bibinfo {pages} {151} (\bibinfo
  {year} {2018})},\ \Eprint {http://arxiv.org/abs/1806.04677} {arXiv:1806.04677
  [hep-ph]} \BibitemShut {NoStop}%
\bibitem [{\citenamefont {Kawasaki}\ \emph {et~al.}(2018)\citenamefont
  {Kawasaki}, \citenamefont {Sekiguchi}, \citenamefont {Yamaguchi},\ and\
  \citenamefont {Yokoyama}}]{Kawasaki:2018bzv}%
  \BibitemOpen
  \bibfield  {author} {\bibinfo {author} {\bibfnamefont {M.}~\bibnamefont
  {Kawasaki}}, \bibinfo {author} {\bibfnamefont {T.}~\bibnamefont {Sekiguchi}},
  \bibinfo {author} {\bibfnamefont {M.}~\bibnamefont {Yamaguchi}}, \ and\
  \bibinfo {author} {\bibfnamefont {J.}~\bibnamefont {Yokoyama}},\ }\href
  {\doibase 10.1093/ptep/pty098} {\bibfield  {journal} {\bibinfo  {journal}
  {PTEP}\ }\textbf {\bibinfo {volume} {2018}},\ \bibinfo {pages} {091E01}
  (\bibinfo {year} {2018})},\ \Eprint {http://arxiv.org/abs/1806.05566}
  {arXiv:1806.05566 [hep-ph]} \BibitemShut {NoStop}%
\bibitem [{\citenamefont {Vaquero}\ \emph {et~al.}(2018)\citenamefont
  {Vaquero}, \citenamefont {Redondo},\ and\ \citenamefont
  {Stadler}}]{Vaquero:2018tib}%
  \BibitemOpen
  \bibfield  {author} {\bibinfo {author} {\bibfnamefont {A.}~\bibnamefont
  {Vaquero}}, \bibinfo {author} {\bibfnamefont {J.}~\bibnamefont {Redondo}}, \
  and\ \bibinfo {author} {\bibfnamefont {J.}~\bibnamefont {Stadler}},\ }\href
  {\doibase 10.1088/1475-7516/2019/04/012} {\  (\bibinfo {year} {2018}),\
  10.1088/1475-7516/2019/04/012},\ \bibinfo {note} {[JCAP1904,012(2019)]},\
  \Eprint {http://arxiv.org/abs/1809.09241} {arXiv:1809.09241 [astro-ph.CO]}
  \BibitemShut {NoStop}%
\bibitem [{\citenamefont {Wilczek}(2004)}]{Wilczek:2004cr}%
  \BibitemOpen
  \bibfield  {author} {\bibinfo {author} {\bibfnamefont {F.}~\bibnamefont
  {Wilczek}},\ }\href@noop {} {\ ,\ \bibinfo {pages} {151} (\bibinfo {year}
  {2004})},\ \Eprint {http://arxiv.org/abs/hep-ph/0408167}
  {arXiv:hep-ph/0408167 [hep-ph]} \BibitemShut {NoStop}%
\bibitem [{\citenamefont {Hertzberg}\ \emph {et~al.}(2008)\citenamefont
  {Hertzberg}, \citenamefont {Tegmark},\ and\ \citenamefont
  {Wilczek}}]{Hertzberg:2008wr}%
  \BibitemOpen
  \bibfield  {author} {\bibinfo {author} {\bibfnamefont {M.~P.}\ \bibnamefont
  {Hertzberg}}, \bibinfo {author} {\bibfnamefont {M.}~\bibnamefont {Tegmark}},
  \ and\ \bibinfo {author} {\bibfnamefont {F.}~\bibnamefont {Wilczek}},\ }\href
  {\doibase 10.1103/PhysRevD.78.083507} {\bibfield  {journal} {\bibinfo
  {journal} {Phys. Rev.}\ }\textbf {\bibinfo {volume} {D78}},\ \bibinfo {pages}
  {083507} (\bibinfo {year} {2008})},\ \Eprint {http://arxiv.org/abs/0807.1726}
  {arXiv:0807.1726 [astro-ph]} \BibitemShut {NoStop}%
\bibitem [{\citenamefont {Freivogel}(2010)}]{Freivogel:2008qc}%
  \BibitemOpen
  \bibfield  {author} {\bibinfo {author} {\bibfnamefont {B.}~\bibnamefont
  {Freivogel}},\ }\href {\doibase 10.1088/1475-7516/2010/03/021} {\bibfield
  {journal} {\bibinfo  {journal} {JCAP}\ }\textbf {\bibinfo {volume} {1003}},\
  \bibinfo {pages} {021} (\bibinfo {year} {2010})},\ \Eprint
  {http://arxiv.org/abs/0810.0703} {arXiv:0810.0703 [hep-th]} \BibitemShut
  {NoStop}%
\bibitem [{\citenamefont {Visinelli}\ and\ \citenamefont
  {Gondolo}(2009)}]{Visinelli:2009zm}%
  \BibitemOpen
  \bibfield  {author} {\bibinfo {author} {\bibfnamefont {L.}~\bibnamefont
  {Visinelli}}\ and\ \bibinfo {author} {\bibfnamefont {P.}~\bibnamefont
  {Gondolo}},\ }\href {\doibase 10.1103/PhysRevD.80.035024} {\bibfield
  {journal} {\bibinfo  {journal} {Phys. Rev.}\ }\textbf {\bibinfo {volume}
  {D80}},\ \bibinfo {pages} {035024} (\bibinfo {year} {2009})},\ \Eprint
  {http://arxiv.org/abs/0903.4377} {arXiv:0903.4377 [astro-ph.CO]} \BibitemShut
  {NoStop}%
\bibitem [{\citenamefont {Hamann}\ \emph {et~al.}(2009)\citenamefont {Hamann},
  \citenamefont {Hannestad}, \citenamefont {Raffelt},\ and\ \citenamefont
  {Wong}}]{Hamann:2009yf}%
  \BibitemOpen
  \bibfield  {author} {\bibinfo {author} {\bibfnamefont {J.}~\bibnamefont
  {Hamann}}, \bibinfo {author} {\bibfnamefont {S.}~\bibnamefont {Hannestad}},
  \bibinfo {author} {\bibfnamefont {G.~G.}\ \bibnamefont {Raffelt}}, \ and\
  \bibinfo {author} {\bibfnamefont {Y.~Y.~Y.}\ \bibnamefont {Wong}},\ }\href
  {\doibase 10.1088/1475-7516/2009/06/022} {\bibfield  {journal} {\bibinfo
  {journal} {JCAP}\ }\textbf {\bibinfo {volume} {0906}},\ \bibinfo {pages}
  {022} (\bibinfo {year} {2009})},\ \Eprint {http://arxiv.org/abs/0904.0647}
  {arXiv:0904.0647 [hep-ph]} \BibitemShut {NoStop}%
\bibitem [{\citenamefont {Hoof}\ \emph {et~al.}(2019)\citenamefont {Hoof},
  \citenamefont {Kahlhoefer}, \citenamefont {Scott}, \citenamefont {Weniger},\
  and\ \citenamefont {White}}]{Hoof:2018ieb}%
  \BibitemOpen
  \bibfield  {author} {\bibinfo {author} {\bibfnamefont {S.}~\bibnamefont
  {Hoof}}, \bibinfo {author} {\bibfnamefont {F.}~\bibnamefont {Kahlhoefer}},
  \bibinfo {author} {\bibfnamefont {P.}~\bibnamefont {Scott}}, \bibinfo
  {author} {\bibfnamefont {C.}~\bibnamefont {Weniger}}, \ and\ \bibinfo
  {author} {\bibfnamefont {M.}~\bibnamefont {White}},\ }\href {\doibase
  10.1007/JHEP11(2019)099, 10.1007/JHEP03(2019)191} {\bibfield  {journal}
  {\bibinfo  {journal} {JHEP}\ }\textbf {\bibinfo {volume} {03}},\ \bibinfo
  {pages} {191} (\bibinfo {year} {2019})},\ \bibinfo {note} {[Erratum:
  JHEP11,099(2019)]},\ \Eprint {http://arxiv.org/abs/1810.07192}
  {arXiv:1810.07192 [hep-ph]} \BibitemShut {NoStop}%
\bibitem [{\citenamefont {Asztalos}\ \emph {et~al.}(2010)\citenamefont
  {Asztalos} \emph {et~al.}}]{Asztalos:2009yp}%
  \BibitemOpen
  \bibfield  {author} {\bibinfo {author} {\bibfnamefont {S.~J.}\ \bibnamefont
  {Asztalos}} \emph {et~al.} (\bibinfo {collaboration} {ADMX}),\ }\href
  {\doibase 10.1103/PhysRevLett.104.041301} {\bibfield  {journal} {\bibinfo
  {journal} {Phys. Rev. Lett.}\ }\textbf {\bibinfo {volume} {104}},\ \bibinfo
  {pages} {041301} (\bibinfo {year} {2010})},\ \Eprint
  {http://arxiv.org/abs/0910.5914} {arXiv:0910.5914 [astro-ph.CO]} \BibitemShut
  {NoStop}%
\bibitem [{\citenamefont {Silva-Feaver}\ \emph {et~al.}(2017)\citenamefont
  {Silva-Feaver} \emph {et~al.}}]{Silva-Feaver:2016qhh}%
  \BibitemOpen
  \bibfield  {author} {\bibinfo {author} {\bibfnamefont {M.}~\bibnamefont
  {Silva-Feaver}} \emph {et~al.},\ }\bibfield  {booktitle} {\emph {\bibinfo
  {booktitle} {{Proceedings, Applied Superconductivity Conference (ASC 2016):
  Denver, Colorado, September 4-9, 2016}}},\ }\href {\doibase
  10.1109/TASC.2016.2631425} {\bibfield  {journal} {\bibinfo  {journal} {IEEE
  Trans. Appl. Supercond.}\ }\textbf {\bibinfo {volume} {27}},\ \bibinfo
  {pages} {1400204} (\bibinfo {year} {2017})},\ \Eprint
  {http://arxiv.org/abs/1610.09344} {arXiv:1610.09344 [astro-ph.IM]}
  \BibitemShut {NoStop}%
\bibitem [{\citenamefont {Caldwell}\ \emph {et~al.}(2017)\citenamefont
  {Caldwell}, \citenamefont {Dvali}, \citenamefont {Majorovits}, \citenamefont
  {Millar}, \citenamefont {Raffelt}, \citenamefont {Redondo}, \citenamefont
  {Reimann}, \citenamefont {Simon},\ and\ \citenamefont
  {Steffen}}]{TheMADMAXWorkingGroup:2016hpc}%
  \BibitemOpen
  \bibfield  {author} {\bibinfo {author} {\bibfnamefont {A.}~\bibnamefont
  {Caldwell}}, \bibinfo {author} {\bibfnamefont {G.}~\bibnamefont {Dvali}},
  \bibinfo {author} {\bibfnamefont {B.}~\bibnamefont {Majorovits}}, \bibinfo
  {author} {\bibfnamefont {A.}~\bibnamefont {Millar}}, \bibinfo {author}
  {\bibfnamefont {G.}~\bibnamefont {Raffelt}}, \bibinfo {author} {\bibfnamefont
  {J.}~\bibnamefont {Redondo}}, \bibinfo {author} {\bibfnamefont
  {O.}~\bibnamefont {Reimann}}, \bibinfo {author} {\bibfnamefont
  {F.}~\bibnamefont {Simon}}, \ and\ \bibinfo {author} {\bibfnamefont
  {F.}~\bibnamefont {Steffen}} (\bibinfo {collaboration} {MADMAX Working
  Group}),\ }\href {\doibase 10.1103/PhysRevLett.118.091801} {\bibfield
  {journal} {\bibinfo  {journal} {Phys. Rev. Lett.}\ }\textbf {\bibinfo
  {volume} {118}},\ \bibinfo {pages} {091801} (\bibinfo {year} {2017})},\
  \Eprint {http://arxiv.org/abs/1611.05865} {arXiv:1611.05865
  [physics.ins-det]} \BibitemShut {NoStop}%
\bibitem [{\citenamefont {Majorovits}\ \emph {et~al.}(2017)\citenamefont
  {Majorovits} \emph {et~al.}}]{Majorovits:2017ppy}%
  \BibitemOpen
  \bibfield  {author} {\bibinfo {author} {\bibfnamefont {B.}~\bibnamefont
  {Majorovits}} \emph {et~al.} (\bibinfo {collaboration} {MADMAX interest
  Group}),\ }in\ \href@noop {} {\emph {\bibinfo {booktitle} {{15th
  International Conference on Topics in Astroparticle and Underground Physics
  (TAUP 2017) Sudbury, Ontario, Canada, July 24-28, 2017}}}}\ (\bibinfo {year}
  {2017})\ \Eprint {http://arxiv.org/abs/1712.01062} {arXiv:1712.01062
  [physics.ins-det]} \BibitemShut {NoStop}%
\bibitem [{\citenamefont {Brun}\ \emph {et~al.}(2019)\citenamefont {Brun} \emph
  {et~al.}}]{Brun:2019lyf}%
  \BibitemOpen
  \bibfield  {author} {\bibinfo {author} {\bibfnamefont {P.}~\bibnamefont
  {Brun}} \emph {et~al.} (\bibinfo {collaboration} {MADMAX}),\ }\href {\doibase
  10.1140/epjc/s10052-019-6683-x} {\bibfield  {journal} {\bibinfo  {journal}
  {Eur. Phys. J.}\ }\textbf {\bibinfo {volume} {C79}},\ \bibinfo {pages} {186}
  (\bibinfo {year} {2019})},\ \Eprint {http://arxiv.org/abs/1901.07401}
  {arXiv:1901.07401 [physics.ins-det]} \BibitemShut {NoStop}%
\bibitem [{\citenamefont {Jackson~Kimball}\ \emph {et~al.}(2017)\citenamefont
  {Jackson~Kimball} \emph {et~al.}}]{JacksonKimball:2017elr}%
  \BibitemOpen
  \bibfield  {author} {\bibinfo {author} {\bibfnamefont {D.~F.}\ \bibnamefont
  {Jackson~Kimball}} \emph {et~al.},\ }\href@noop {} {\  (\bibinfo {year}
  {2017})},\ \Eprint {http://arxiv.org/abs/1711.08999} {arXiv:1711.08999
  [physics.ins-det]} \BibitemShut {NoStop}%
\bibitem [{\citenamefont {Anastassopoulos}\ \emph {et~al.}(2017)\citenamefont
  {Anastassopoulos} \emph {et~al.}}]{Anastassopoulos:2017ftl}%
  \BibitemOpen
  \bibfield  {author} {\bibinfo {author} {\bibfnamefont {V.}~\bibnamefont
  {Anastassopoulos}} \emph {et~al.} (\bibinfo {collaboration} {CAST}),\ }\href
  {\doibase 10.1038/nphys4109} {\bibfield  {journal} {\bibinfo  {journal}
  {Nature Phys.}\ }\textbf {\bibinfo {volume} {13}},\ \bibinfo {pages} {584}
  (\bibinfo {year} {2017})},\ \Eprint {http://arxiv.org/abs/1705.02290}
  {arXiv:1705.02290 [hep-ex]} \BibitemShut {NoStop}%
\bibitem [{\citenamefont {Zhong}\ \emph {et~al.}(2018)\citenamefont {Zhong}
  \emph {et~al.}}]{Zhong:2018rsr}%
  \BibitemOpen
  \bibfield  {author} {\bibinfo {author} {\bibfnamefont {L.}~\bibnamefont
  {Zhong}} \emph {et~al.} (\bibinfo {collaboration} {HAYSTAC}),\ }\href
  {\doibase 10.1103/PhysRevD.97.092001} {\bibfield  {journal} {\bibinfo
  {journal} {Phys. Rev.}\ }\textbf {\bibinfo {volume} {D97}},\ \bibinfo {pages}
  {092001} (\bibinfo {year} {2018})},\ \Eprint
  {http://arxiv.org/abs/1803.03690} {arXiv:1803.03690 [hep-ex]} \BibitemShut
  {NoStop}%
\bibitem [{\citenamefont {Du}\ \emph {et~al.}(2018)\citenamefont {Du} \emph
  {et~al.}}]{Du:2018uak}%
  \BibitemOpen
  \bibfield  {author} {\bibinfo {author} {\bibfnamefont {N.}~\bibnamefont {Du}}
  \emph {et~al.} (\bibinfo {collaboration} {ADMX}),\ }\href {\doibase
  10.1103/PhysRevLett.120.151301} {\bibfield  {journal} {\bibinfo  {journal}
  {Phys. Rev. Lett.}\ }\textbf {\bibinfo {volume} {120}},\ \bibinfo {pages}
  {151301} (\bibinfo {year} {2018})},\ \Eprint
  {http://arxiv.org/abs/1804.05750} {arXiv:1804.05750 [hep-ex]} \BibitemShut
  {NoStop}%
\bibitem [{\citenamefont {Ouellet}\ \emph {et~al.}(2019)\citenamefont {Ouellet}
  \emph {et~al.}}]{Ouellet:2018beu}%
  \BibitemOpen
  \bibfield  {author} {\bibinfo {author} {\bibfnamefont {J.~L.}\ \bibnamefont
  {Ouellet}} \emph {et~al.},\ }\href {\doibase 10.1103/PhysRevLett.122.121802}
  {\bibfield  {journal} {\bibinfo  {journal} {Phys. Rev. Lett.}\ }\textbf
  {\bibinfo {volume} {122}},\ \bibinfo {pages} {121802} (\bibinfo {year}
  {2019})},\ \Eprint {http://arxiv.org/abs/1810.12257} {arXiv:1810.12257
  [hep-ex]} \BibitemShut {NoStop}%
\bibitem [{\citenamefont {Shokair}\ \emph {et~al.}(2014)\citenamefont {Shokair}
  \emph {et~al.}}]{Shokair:2014rna}%
  \BibitemOpen
  \bibfield  {author} {\bibinfo {author} {\bibfnamefont {T.~M.}\ \bibnamefont
  {Shokair}} \emph {et~al.},\ }\href {\doibase 10.1142/S0217751X14430040}
  {\bibfield  {journal} {\bibinfo  {journal} {Int. J. Mod. Phys.}\ }\textbf
  {\bibinfo {volume} {A29}},\ \bibinfo {pages} {1443004} (\bibinfo {year}
  {2014})},\ \Eprint {http://arxiv.org/abs/1405.3685} {arXiv:1405.3685
  [physics.ins-det]} \BibitemShut {NoStop}%
\bibitem [{\citenamefont {Al~Kenany}\ \emph {et~al.}(2017)\citenamefont
  {Al~Kenany} \emph {et~al.}}]{Kenany:2016tta}%
  \BibitemOpen
  \bibfield  {author} {\bibinfo {author} {\bibfnamefont {S.}~\bibnamefont
  {Al~Kenany}} \emph {et~al.},\ }\href {\doibase 10.1016/j.nima.2017.02.012}
  {\bibfield  {journal} {\bibinfo  {journal} {Nucl. Instrum. Meth.}\ }\textbf
  {\bibinfo {volume} {A854}},\ \bibinfo {pages} {11} (\bibinfo {year}
  {2017})},\ \Eprint {http://arxiv.org/abs/1611.07123} {arXiv:1611.07123
  [physics.ins-det]} \BibitemShut {NoStop}%
\bibitem [{\citenamefont {Brubaker}\ \emph {et~al.}(2017)\citenamefont
  {Brubaker} \emph {et~al.}}]{Brubaker:2016ktl}%
  \BibitemOpen
  \bibfield  {author} {\bibinfo {author} {\bibfnamefont {B.~M.}\ \bibnamefont
  {Brubaker}} \emph {et~al.},\ }\href {\doibase 10.1103/PhysRevLett.118.061302}
  {\bibfield  {journal} {\bibinfo  {journal} {Phys. Rev. Lett.}\ }\textbf
  {\bibinfo {volume} {118}},\ \bibinfo {pages} {061302} (\bibinfo {year}
  {2017})},\ \Eprint {http://arxiv.org/abs/1610.02580} {arXiv:1610.02580
  [astro-ph.CO]} \BibitemShut {NoStop}%
\bibitem [{\citenamefont {Kahn}\ \emph {et~al.}(2016)\citenamefont {Kahn},
  \citenamefont {Safdi},\ and\ \citenamefont {Thaler}}]{Kahn:2016aff}%
  \BibitemOpen
  \bibfield  {author} {\bibinfo {author} {\bibfnamefont {Y.}~\bibnamefont
  {Kahn}}, \bibinfo {author} {\bibfnamefont {B.~R.}\ \bibnamefont {Safdi}}, \
  and\ \bibinfo {author} {\bibfnamefont {J.}~\bibnamefont {Thaler}},\ }\href
  {\doibase 10.1103/PhysRevLett.117.141801} {\bibfield  {journal} {\bibinfo
  {journal} {Phys. Rev. Lett.}\ }\textbf {\bibinfo {volume} {117}},\ \bibinfo
  {pages} {141801} (\bibinfo {year} {2016})},\ \Eprint
  {http://arxiv.org/abs/1602.01086} {arXiv:1602.01086 [hep-ph]} \BibitemShut
  {NoStop}%
\bibitem [{\citenamefont {McAllister}\ \emph {et~al.}(2017)\citenamefont
  {McAllister}, \citenamefont {Flower}, \citenamefont {Ivanov}, \citenamefont
  {Goryachev}, \citenamefont {Bourhill},\ and\ \citenamefont
  {Tobar}}]{McAllister:2017lkb}%
  \BibitemOpen
  \bibfield  {author} {\bibinfo {author} {\bibfnamefont {B.~T.}\ \bibnamefont
  {McAllister}}, \bibinfo {author} {\bibfnamefont {G.}~\bibnamefont {Flower}},
  \bibinfo {author} {\bibfnamefont {E.~N.}\ \bibnamefont {Ivanov}}, \bibinfo
  {author} {\bibfnamefont {M.}~\bibnamefont {Goryachev}}, \bibinfo {author}
  {\bibfnamefont {J.}~\bibnamefont {Bourhill}}, \ and\ \bibinfo {author}
  {\bibfnamefont {M.~E.}\ \bibnamefont {Tobar}},\ }\href {\doibase
  10.1016/j.dark.2017.09.010} {\bibfield  {journal} {\bibinfo  {journal} {Phys.
  Dark Univ.}\ }\textbf {\bibinfo {volume} {18}},\ \bibinfo {pages} {67}
  (\bibinfo {year} {2017})},\ \Eprint {http://arxiv.org/abs/1706.00209}
  {arXiv:1706.00209 [physics.ins-det]} \BibitemShut {NoStop}%
\bibitem [{\citenamefont {Alesini}\ \emph {et~al.}(2017)\citenamefont
  {Alesini}, \citenamefont {Babusci}, \citenamefont {Di~Gioacchino},
  \citenamefont {Gatti}, \citenamefont {Lamanna},\ and\ \citenamefont
  {Ligi}}]{Alesini:2017ifp}%
  \BibitemOpen
  \bibfield  {author} {\bibinfo {author} {\bibfnamefont {D.}~\bibnamefont
  {Alesini}}, \bibinfo {author} {\bibfnamefont {D.}~\bibnamefont {Babusci}},
  \bibinfo {author} {\bibfnamefont {D.}~\bibnamefont {Di~Gioacchino}}, \bibinfo
  {author} {\bibfnamefont {C.}~\bibnamefont {Gatti}}, \bibinfo {author}
  {\bibfnamefont {G.}~\bibnamefont {Lamanna}}, \ and\ \bibinfo {author}
  {\bibfnamefont {C.}~\bibnamefont {Ligi}},\ }\href@noop {} {\  (\bibinfo
  {year} {2017})},\ \Eprint {http://arxiv.org/abs/1707.06010} {arXiv:1707.06010
  [physics.ins-det]} \BibitemShut {NoStop}%
\bibitem [{\citenamefont {Lawson}\ \emph {et~al.}(2019)\citenamefont {Lawson},
  \citenamefont {Millar}, \citenamefont {Pancaldi}, \citenamefont
  {Vitagliano},\ and\ \citenamefont {Wilczek}}]{Lawson:2019brd}%
  \BibitemOpen
  \bibfield  {author} {\bibinfo {author} {\bibfnamefont {M.}~\bibnamefont
  {Lawson}}, \bibinfo {author} {\bibfnamefont {A.~J.}\ \bibnamefont {Millar}},
  \bibinfo {author} {\bibfnamefont {M.}~\bibnamefont {Pancaldi}}, \bibinfo
  {author} {\bibfnamefont {E.}~\bibnamefont {Vitagliano}}, \ and\ \bibinfo
  {author} {\bibfnamefont {F.}~\bibnamefont {Wilczek}},\ }\href {\doibase
  10.1103/PhysRevLett.123.141802} {\bibfield  {journal} {\bibinfo  {journal}
  {Phys. Rev. Lett.}\ }\textbf {\bibinfo {volume} {123}},\ \bibinfo {pages}
  {141802} (\bibinfo {year} {2019})},\ \Eprint
  {http://arxiv.org/abs/1904.11872} {arXiv:1904.11872 [hep-ph]} \BibitemShut
  {NoStop}%
\bibitem [{\citenamefont {Irastorza}\ and\ \citenamefont
  {Redondo}(2018)}]{Irastorza:2018dyq}%
  \BibitemOpen
  \bibfield  {author} {\bibinfo {author} {\bibfnamefont {I.~G.}\ \bibnamefont
  {Irastorza}}\ and\ \bibinfo {author} {\bibfnamefont {J.}~\bibnamefont
  {Redondo}},\ }\href {\doibase 10.1016/j.ppnp.2018.05.003} {\bibfield
  {journal} {\bibinfo  {journal} {Prog. Part. Nucl. Phys.}\ }\textbf {\bibinfo
  {volume} {102}},\ \bibinfo {pages} {89} (\bibinfo {year} {2018})},\ \Eprint
  {http://arxiv.org/abs/1801.08127} {arXiv:1801.08127 [hep-ph]} \BibitemShut
  {NoStop}%
\bibitem [{\citenamefont {Kim}(1979)}]{Kim:1979if}%
  \BibitemOpen
  \bibfield  {author} {\bibinfo {author} {\bibfnamefont {J.~E.}\ \bibnamefont
  {Kim}},\ }\href {\doibase 10.1103/PhysRevLett.43.103} {\bibfield  {journal}
  {\bibinfo  {journal} {Phys. Rev. Lett.}\ }\textbf {\bibinfo {volume} {43}},\
  \bibinfo {pages} {103} (\bibinfo {year} {1979})}\BibitemShut {NoStop}%
\bibitem [{\citenamefont {Shifman}\ \emph {et~al.}(1980)\citenamefont
  {Shifman}, \citenamefont {Vainshtein},\ and\ \citenamefont
  {Zakharov}}]{Shifman:1979if}%
  \BibitemOpen
  \bibfield  {author} {\bibinfo {author} {\bibfnamefont {M.~A.}\ \bibnamefont
  {Shifman}}, \bibinfo {author} {\bibfnamefont {A.~I.}\ \bibnamefont
  {Vainshtein}}, \ and\ \bibinfo {author} {\bibfnamefont {V.~I.}\ \bibnamefont
  {Zakharov}},\ }\href {\doibase 10.1016/0550-3213(80)90209-6} {\bibfield
  {journal} {\bibinfo  {journal} {Nucl. Phys.}\ }\textbf {\bibinfo {volume}
  {B166}},\ \bibinfo {pages} {493} (\bibinfo {year} {1980})}\BibitemShut
  {NoStop}%
\bibitem [{\citenamefont {Zhitnitsky}(1980)}]{Zhitnitsky:1980tq}%
  \BibitemOpen
  \bibfield  {author} {\bibinfo {author} {\bibfnamefont {A.~R.}\ \bibnamefont
  {Zhitnitsky}},\ }\href@noop {} {\bibfield  {journal} {\bibinfo  {journal}
  {Sov. J. Nucl. Phys.}\ }\textbf {\bibinfo {volume} {31}},\ \bibinfo {pages}
  {260} (\bibinfo {year} {1980})},\ \bibinfo {note} {[Yad.
  Fiz.31,497(1980)]}\BibitemShut {NoStop}%
\bibitem [{\citenamefont {Dine}\ \emph {et~al.}(1981)\citenamefont {Dine},
  \citenamefont {Fischler},\ and\ \citenamefont {Srednicki}}]{Dine:1981rt}%
  \BibitemOpen
  \bibfield  {author} {\bibinfo {author} {\bibfnamefont {M.}~\bibnamefont
  {Dine}}, \bibinfo {author} {\bibfnamefont {W.}~\bibnamefont {Fischler}}, \
  and\ \bibinfo {author} {\bibfnamefont {M.}~\bibnamefont {Srednicki}},\ }\href
  {\doibase 10.1016/0370-2693(81)90590-6} {\bibfield  {journal} {\bibinfo
  {journal} {Phys. Lett.}\ }\textbf {\bibinfo {volume} {104B}},\ \bibinfo
  {pages} {199} (\bibinfo {year} {1981})}\BibitemShut {NoStop}%
\bibitem [{\citenamefont {Conlon}(2007)}]{Conlon:2006gv}%
  \BibitemOpen
  \bibfield  {author} {\bibinfo {author} {\bibfnamefont {J.~P.}\ \bibnamefont
  {Conlon}},\ }\href {\doibase 10.1002/prop.200610334} {\bibfield  {journal}
  {\bibinfo  {journal} {Fortsch. Phys.}\ }\textbf {\bibinfo {volume} {55}},\
  \bibinfo {pages} {287} (\bibinfo {year} {2007})},\ \Eprint
  {http://arxiv.org/abs/hep-th/0611039} {arXiv:hep-th/0611039 [hep-th]}
  \BibitemShut {NoStop}%
\bibitem [{\citenamefont {Grana}(2006)}]{Grana:2005jc}%
  \BibitemOpen
  \bibfield  {author} {\bibinfo {author} {\bibfnamefont {M.}~\bibnamefont
  {Grana}},\ }\href {\doibase 10.1016/j.physrep.2005.10.008} {\bibfield
  {journal} {\bibinfo  {journal} {Phys. Rept.}\ }\textbf {\bibinfo {volume}
  {423}},\ \bibinfo {pages} {91} (\bibinfo {year} {2006})},\ \Eprint
  {http://arxiv.org/abs/hep-th/0509003} {arXiv:hep-th/0509003 [hep-th]}
  \BibitemShut {NoStop}%
\bibitem [{\citenamefont {Caputo}\ \emph {et~al.}(2018)\citenamefont {Caputo},
  \citenamefont {Garay},\ and\ \citenamefont {Witte}}]{Caputo:2018ljp}%
  \BibitemOpen
  \bibfield  {author} {\bibinfo {author} {\bibfnamefont {A.}~\bibnamefont
  {Caputo}}, \bibinfo {author} {\bibfnamefont {C.~P.}\ \bibnamefont {Garay}}, \
  and\ \bibinfo {author} {\bibfnamefont {S.~J.}\ \bibnamefont {Witte}},\ }\href
  {\doibase 10.1103/PhysRevD.99.089901, 10.1103/PhysRevD.98.083024} {\bibfield
  {journal} {\bibinfo  {journal} {Phys. Rev.}\ }\textbf {\bibinfo {volume}
  {D98}},\ \bibinfo {pages} {083024} (\bibinfo {year} {2018})},\ \bibinfo
  {note} {[Erratum: Phys. Rev.D99,no.8,089901(2019)]},\ \Eprint
  {http://arxiv.org/abs/1805.08780} {arXiv:1805.08780 [astro-ph.CO]}
  \BibitemShut {NoStop}%
\bibitem [{\citenamefont {Caputo}\ \emph {et~al.}(2019)\citenamefont {Caputo},
  \citenamefont {Regis}, \citenamefont {Taoso},\ and\ \citenamefont
  {Witte}}]{Caputo:2018vmy}%
  \BibitemOpen
  \bibfield  {author} {\bibinfo {author} {\bibfnamefont {A.}~\bibnamefont
  {Caputo}}, \bibinfo {author} {\bibfnamefont {M.}~\bibnamefont {Regis}},
  \bibinfo {author} {\bibfnamefont {M.}~\bibnamefont {Taoso}}, \ and\ \bibinfo
  {author} {\bibfnamefont {S.~J.}\ \bibnamefont {Witte}},\ }\href {\doibase
  10.1088/1475-7516/2019/03/027} {\bibfield  {journal} {\bibinfo  {journal}
  {JCAP}\ }\textbf {\bibinfo {volume} {1903}},\ \bibinfo {pages} {027}
  (\bibinfo {year} {2019})},\ \Eprint {http://arxiv.org/abs/1811.08436}
  {arXiv:1811.08436 [hep-ph]} \BibitemShut {NoStop}%
\bibitem [{\citenamefont {Carenza}\ \emph {et~al.}(2019)\citenamefont
  {Carenza}, \citenamefont {Mirizzi},\ and\ \citenamefont
  {Sigl}}]{Carenza:2019vzg}%
  \BibitemOpen
  \bibfield  {author} {\bibinfo {author} {\bibfnamefont {P.}~\bibnamefont
  {Carenza}}, \bibinfo {author} {\bibfnamefont {A.}~\bibnamefont {Mirizzi}}, \
  and\ \bibinfo {author} {\bibfnamefont {G.}~\bibnamefont {Sigl}},\ }\href@noop
  {} {\  (\bibinfo {year} {2019})},\ \Eprint {http://arxiv.org/abs/1911.07838}
  {arXiv:1911.07838 [hep-ph]} \BibitemShut {NoStop}%
\bibitem [{\citenamefont {Raffelt}\ and\ \citenamefont
  {Dearborn}(1987)}]{Raffelt:1987yu}%
  \BibitemOpen
  \bibfield  {author} {\bibinfo {author} {\bibfnamefont {G.~G.}\ \bibnamefont
  {Raffelt}}\ and\ \bibinfo {author} {\bibfnamefont {D.~S.~P.}\ \bibnamefont
  {Dearborn}},\ }\href {\doibase 10.1103/PhysRevD.36.2211} {\bibfield
  {journal} {\bibinfo  {journal} {Phys. Rev.}\ }\textbf {\bibinfo {volume}
  {D36}},\ \bibinfo {pages} {2211} (\bibinfo {year} {1987})}\BibitemShut
  {NoStop}%
\bibitem [{\citenamefont {Raffelt}(2008)}]{Raffelt:2006cw}%
  \BibitemOpen
  \bibfield  {author} {\bibinfo {author} {\bibfnamefont {G.~G.}\ \bibnamefont
  {Raffelt}},\ }\bibfield  {booktitle} {\emph {\bibinfo {booktitle} {{Axions:
  Theory, cosmology, and experimental searches. Proceedings, 1st Joint
  ILIAS-CERN-CAST axion training, Geneva, Switzerland, November 30-December 2,
  2005}}},\ }\href {\doibase 10.1007/978-3-540-73518-2_3} {\bibfield  {journal}
  {\bibinfo  {journal} {Lect. Notes Phys.}\ }\textbf {\bibinfo {volume}
  {741}},\ \bibinfo {pages} {51} (\bibinfo {year} {2008})},\ \bibinfo {note}
  {[,51(2006)]},\ \Eprint {http://arxiv.org/abs/hep-ph/0611350}
  {arXiv:hep-ph/0611350 [hep-ph]} \BibitemShut {NoStop}%
\bibitem [{\citenamefont {Friedland}\ \emph {et~al.}(2013)\citenamefont
  {Friedland}, \citenamefont {Giannotti},\ and\ \citenamefont
  {Wise}}]{Friedland:2012hj}%
  \BibitemOpen
  \bibfield  {author} {\bibinfo {author} {\bibfnamefont {A.}~\bibnamefont
  {Friedland}}, \bibinfo {author} {\bibfnamefont {M.}~\bibnamefont
  {Giannotti}}, \ and\ \bibinfo {author} {\bibfnamefont {M.}~\bibnamefont
  {Wise}},\ }\href {\doibase 10.1103/PhysRevLett.110.061101} {\bibfield
  {journal} {\bibinfo  {journal} {Phys. Rev. Lett.}\ }\textbf {\bibinfo
  {volume} {110}},\ \bibinfo {pages} {061101} (\bibinfo {year} {2013})},\
  \Eprint {http://arxiv.org/abs/1210.1271} {arXiv:1210.1271 [hep-ph]}
  \BibitemShut {NoStop}%
\bibitem [{\citenamefont {Hook}\ \emph {et~al.}(2018)\citenamefont {Hook},
  \citenamefont {Kahn}, \citenamefont {Safdi},\ and\ \citenamefont
  {Sun}}]{Hook:2018iia}%
  \BibitemOpen
  \bibfield  {author} {\bibinfo {author} {\bibfnamefont {A.}~\bibnamefont
  {Hook}}, \bibinfo {author} {\bibfnamefont {Y.}~\bibnamefont {Kahn}}, \bibinfo
  {author} {\bibfnamefont {B.~R.}\ \bibnamefont {Safdi}}, \ and\ \bibinfo
  {author} {\bibfnamefont {Z.}~\bibnamefont {Sun}},\ }\href {\doibase
  10.1103/PhysRevLett.121.241102} {\bibfield  {journal} {\bibinfo  {journal}
  {Phys. Rev. Lett.}\ }\textbf {\bibinfo {volume} {121}},\ \bibinfo {pages}
  {241102} (\bibinfo {year} {2018})},\ \Eprint
  {http://arxiv.org/abs/1804.03145} {arXiv:1804.03145 [hep-ph]} \BibitemShut
  {NoStop}%
\bibitem [{\citenamefont {Pshirkov}\ and\ \citenamefont
  {Popov}(2009)}]{Pshirkov:2007st}%
  \BibitemOpen
  \bibfield  {author} {\bibinfo {author} {\bibfnamefont {M.~S.}\ \bibnamefont
  {Pshirkov}}\ and\ \bibinfo {author} {\bibfnamefont {S.~B.}\ \bibnamefont
  {Popov}},\ }\href {\doibase 10.1134/S1063776109030030} {\bibfield  {journal}
  {\bibinfo  {journal} {J. Exp. Theor. Phys.}\ }\textbf {\bibinfo {volume}
  {108}},\ \bibinfo {pages} {384} (\bibinfo {year} {2009})},\ \Eprint
  {http://arxiv.org/abs/0711.1264} {arXiv:0711.1264 [astro-ph]} \BibitemShut
  {NoStop}%
\bibitem [{\citenamefont {Huang}\ \emph {et~al.}(2018)\citenamefont {Huang},
  \citenamefont {Kadota}, \citenamefont {Sekiguchi},\ and\ \citenamefont
  {Tashiro}}]{Huang:2018lxq}%
  \BibitemOpen
  \bibfield  {author} {\bibinfo {author} {\bibfnamefont {F.~P.}\ \bibnamefont
  {Huang}}, \bibinfo {author} {\bibfnamefont {K.}~\bibnamefont {Kadota}},
  \bibinfo {author} {\bibfnamefont {T.}~\bibnamefont {Sekiguchi}}, \ and\
  \bibinfo {author} {\bibfnamefont {H.}~\bibnamefont {Tashiro}},\ }\href
  {\doibase 10.1103/PhysRevD.97.123001} {\bibfield  {journal} {\bibinfo
  {journal} {Phys. Rev.}\ }\textbf {\bibinfo {volume} {D97}},\ \bibinfo {pages}
  {123001} (\bibinfo {year} {2018})},\ \Eprint
  {http://arxiv.org/abs/1803.08230} {arXiv:1803.08230 [hep-ph]} \BibitemShut
  {NoStop}%
\bibitem [{\citenamefont {Battye}\ \emph {et~al.}(2019)\citenamefont {Battye},
  \citenamefont {Garbrecht}, \citenamefont {McDonald}, \citenamefont {Pace},\
  and\ \citenamefont {Srinivasan}}]{Battye:2019aco}%
  \BibitemOpen
  \bibfield  {author} {\bibinfo {author} {\bibfnamefont {R.~A.}\ \bibnamefont
  {Battye}}, \bibinfo {author} {\bibfnamefont {B.}~\bibnamefont {Garbrecht}},
  \bibinfo {author} {\bibfnamefont {J.~I.}\ \bibnamefont {McDonald}}, \bibinfo
  {author} {\bibfnamefont {F.}~\bibnamefont {Pace}}, \ and\ \bibinfo {author}
  {\bibfnamefont {S.}~\bibnamefont {Srinivasan}},\ }\href@noop {} {\  (\bibinfo
  {year} {2019})},\ \Eprint {http://arxiv.org/abs/1910.11907} {arXiv:1910.11907
  [astro-ph.CO]} \BibitemShut {NoStop}%
\bibitem [{\citenamefont {Ajello}\ \emph {et~al.}(2017)\citenamefont {Ajello}
  \emph {et~al.}}]{Fermi-LAT:2017yoi}%
  \BibitemOpen
  \bibfield  {author} {\bibinfo {author} {\bibfnamefont {M.}~\bibnamefont
  {Ajello}} \emph {et~al.} (\bibinfo {collaboration} {Fermi-LAT}),\ }\href@noop
  {} {\bibfield  {journal} {\bibinfo  {journal} {Submitted to: Astrophys. J.}\
  } (\bibinfo {year} {2017})},\ \Eprint {http://arxiv.org/abs/1705.00009}
  {arXiv:1705.00009 [astro-ph.HE]} \BibitemShut {NoStop}%
\bibitem [{\citenamefont {Safdi}\ \emph {et~al.}(2019)\citenamefont {Safdi},
  \citenamefont {Sun},\ and\ \citenamefont {Chen}}]{Safdi:2018oeu}%
  \BibitemOpen
  \bibfield  {author} {\bibinfo {author} {\bibfnamefont {B.~R.}\ \bibnamefont
  {Safdi}}, \bibinfo {author} {\bibfnamefont {Z.}~\bibnamefont {Sun}}, \ and\
  \bibinfo {author} {\bibfnamefont {A.~Y.}\ \bibnamefont {Chen}},\ }\href
  {\doibase 10.1103/PhysRevD.99.123021} {\bibfield  {journal} {\bibinfo
  {journal} {Phys. Rev.}\ }\textbf {\bibinfo {volume} {D99}},\ \bibinfo {pages}
  {123021} (\bibinfo {year} {2019})},\ \Eprint
  {http://arxiv.org/abs/1811.01020} {arXiv:1811.01020 [astro-ph.CO]}
  \BibitemShut {NoStop}%
\bibitem [{\citenamefont {Edwards}\ \emph {et~al.}(2019)\citenamefont
  {Edwards}, \citenamefont {Chianese}, \citenamefont {Kavanagh}, \citenamefont
  {Nissanke},\ and\ \citenamefont {Weniger}}]{Edwards:2019tzf}%
  \BibitemOpen
  \bibfield  {author} {\bibinfo {author} {\bibfnamefont {T.~D.~P.}\
  \bibnamefont {Edwards}}, \bibinfo {author} {\bibfnamefont {M.}~\bibnamefont
  {Chianese}}, \bibinfo {author} {\bibfnamefont {B.~J.}\ \bibnamefont
  {Kavanagh}}, \bibinfo {author} {\bibfnamefont {S.~M.}\ \bibnamefont
  {Nissanke}}, \ and\ \bibinfo {author} {\bibfnamefont {C.}~\bibnamefont
  {Weniger}},\ }\href@noop {} {\  (\bibinfo {year} {2019})},\ \Eprint
  {http://arxiv.org/abs/1905.04686} {arXiv:1905.04686 [hep-ph]} \BibitemShut
  {NoStop}%
\bibitem [{\citenamefont {Raffelt}\ and\ \citenamefont
  {Stodolsky}(1988)}]{Raffelt:1987im}%
  \BibitemOpen
  \bibfield  {author} {\bibinfo {author} {\bibfnamefont {G.}~\bibnamefont
  {Raffelt}}\ and\ \bibinfo {author} {\bibfnamefont {L.}~\bibnamefont
  {Stodolsky}},\ }\href {\doibase 10.1103/PhysRevD.37.1237} {\bibfield
  {journal} {\bibinfo  {journal} {Phys.\ Rev.\ D}\ }\textbf {\bibinfo {volume}
  {37}},\ \bibinfo {pages} {1237} (\bibinfo {year} {1988})}\BibitemShut
  {NoStop}%
\bibitem [{\citenamefont {Liouville}(1838)}]{Liouville:1838zza}%
  \BibitemOpen
  \bibfield  {author} {\bibinfo {author} {\bibfnamefont {J.}~\bibnamefont
  {Liouville}},\ }\href {https://gallica.bnf.fr/ark:/12148/bpt6k16382m}
  {\bibfield  {journal} {\bibinfo  {journal} {J. Math. Pure. Appl.}\ }\textbf
  {\bibinfo {volume} {3}},\ \bibinfo {pages} {342} (\bibinfo {year}
  {1838})}\BibitemShut {NoStop}%
\bibitem [{\citenamefont {Piffl}\ \emph {et~al.}(2014)\citenamefont {Piffl}
  \emph {et~al.}}]{Piffl:2013mla}%
  \BibitemOpen
  \bibfield  {author} {\bibinfo {author} {\bibfnamefont {T.}~\bibnamefont
  {Piffl}} \emph {et~al.},\ }\href {\doibase 10.1051/0004-6361/201322531}
  {\bibfield  {journal} {\bibinfo  {journal} {Astron. Astrophys.}\ }\textbf
  {\bibinfo {volume} {562}},\ \bibinfo {pages} {A91} (\bibinfo {year}
  {2014})},\ \Eprint {http://arxiv.org/abs/1309.4293} {arXiv:1309.4293
  [astro-ph.GA]} \BibitemShut {NoStop}%
\bibitem [{\citenamefont {Smith}\ \emph {et~al.}(2007)\citenamefont {Smith}
  \emph {et~al.}}]{Smith:2006ym}%
  \BibitemOpen
  \bibfield  {author} {\bibinfo {author} {\bibfnamefont {M.~C.}\ \bibnamefont
  {Smith}} \emph {et~al.},\ }\href {\doibase 10.1111/j.1365-2966.2007.11964.x}
  {\bibfield  {journal} {\bibinfo  {journal} {Mon. Not. Roy. Astron. Soc.}\
  }\textbf {\bibinfo {volume} {379}},\ \bibinfo {pages} {755} (\bibinfo {year}
  {2007})},\ \Eprint {http://arxiv.org/abs/astro-ph/0611671}
  {arXiv:astro-ph/0611671 [astro-ph]} \BibitemShut {NoStop}%
\bibitem [{\citenamefont {{Alenazi}}\ and\ \citenamefont
  {{Gondolo}}(2006)}]{2006PhRvD..74h3518A}%
  \BibitemOpen
  \bibfield  {author} {\bibinfo {author} {\bibfnamefont {M.~S.}\ \bibnamefont
  {{Alenazi}}}\ and\ \bibinfo {author} {\bibfnamefont {P.}~\bibnamefont
  {{Gondolo}}},\ }\href {\doibase 10.1103/PhysRevD.74.083518} {\bibfield
  {journal} {\bibinfo  {journal} {\prd}\ }\textbf {\bibinfo {volume} {74}},\
  \bibinfo {eid} {083518} (\bibinfo {year} {2006})},\ \Eprint
  {http://arxiv.org/abs/astro-ph/0608390} {arXiv:astro-ph/0608390 [astro-ph]}
  \BibitemShut {NoStop}%
\bibitem [{\citenamefont {{Goldreich}}\ and\ \citenamefont
  {{Julian}}(1969)}]{1969ApJ...157..869G}%
  \BibitemOpen
  \bibfield  {author} {\bibinfo {author} {\bibfnamefont {P.}~\bibnamefont
  {{Goldreich}}}\ and\ \bibinfo {author} {\bibfnamefont {W.~H.}\ \bibnamefont
  {{Julian}}},\ }\href {\doibase 10.1086/150119} {\bibfield  {journal}
  {\bibinfo  {journal} {\apj}\ }\textbf {\bibinfo {volume} {157}},\ \bibinfo
  {pages} {869} (\bibinfo {year} {1969})}\BibitemShut {NoStop}%
\bibitem [{\citenamefont {Pétri}(2016)}]{Petri:2016tqe}%
  \BibitemOpen
  \bibfield  {author} {\bibinfo {author} {\bibfnamefont {J.}~\bibnamefont
  {Pétri}},\ }\href {\doibase 10.1017/S0022377816000763} {\bibfield  {journal}
  {\bibinfo  {journal} {J. Plasma Phys.}\ }\textbf {\bibinfo {volume} {82}},\
  \bibinfo {pages} {635820502} (\bibinfo {year} {2016})},\ \Eprint
  {http://arxiv.org/abs/1608.04895} {arXiv:1608.04895 [astro-ph.HE]}
  \BibitemShut {NoStop}%
\bibitem [{\citenamefont {Carlson}\ and\ \citenamefont
  {Garretson}(1994)}]{Carlson:1994yqa}%
  \BibitemOpen
  \bibfield  {author} {\bibinfo {author} {\bibfnamefont {E.~D.}\ \bibnamefont
  {Carlson}}\ and\ \bibinfo {author} {\bibfnamefont {W.}~\bibnamefont
  {Garretson}},\ }\href {\doibase 10.1016/0370-2693(94)90555-X} {\bibfield
  {journal} {\bibinfo  {journal} {Phys. Lett. B}\ }\textbf {\bibinfo {volume}
  {336}},\ \bibinfo {pages} {431} (\bibinfo {year} {1994})}\BibitemShut
  {NoStop}%
\bibitem [{\citenamefont {{Timokhin}}\ and\ \citenamefont
  {{Harding}}(2019)}]{2019ApJ...871...12T}%
  \BibitemOpen
  \bibfield  {author} {\bibinfo {author} {\bibfnamefont {A.~N.}\ \bibnamefont
  {{Timokhin}}}\ and\ \bibinfo {author} {\bibfnamefont {A.~K.}\ \bibnamefont
  {{Harding}}},\ }\href {\doibase 10.3847/1538-4357/aaf050} {\bibfield
  {journal} {\bibinfo  {journal} {\apj}\ }\textbf {\bibinfo {volume} {871}},\
  \bibinfo {eid} {12} (\bibinfo {year} {2019})},\ \Eprint
  {http://arxiv.org/abs/1803.08924} {arXiv:1803.08924 [astro-ph.HE]}
  \BibitemShut {NoStop}%
\bibitem [{\citenamefont {Krause-Polstorff}\ and\ \citenamefont
  {Michel}(1985)}]{10.1093/mnras/213.1.43P}%
  \BibitemOpen
  \bibfield  {author} {\bibinfo {author} {\bibfnamefont {J.}~\bibnamefont
  {Krause-Polstorff}}\ and\ \bibinfo {author} {\bibfnamefont {F.~C.}\
  \bibnamefont {Michel}},\ }\href {\doibase 10.1093/mnras/213.1.43P} {\bibfield
   {journal} {\bibinfo  {journal} {Monthly Notices of the Royal Astronomical
  Society}\ }\textbf {\bibinfo {volume} {213}},\ \bibinfo {pages} {43P}
  (\bibinfo {year} {1985})},\ \Eprint
  {http://arxiv.org/abs/http://oup.prod.sis.lan/mnras/article-pdf/213/1/43P/2966415/mnras213-043P.pdf}
  {http://oup.prod.sis.lan/mnras/article-pdf/213/1/43P/2966415/mnras213-043P.pdf}
  \BibitemShut {NoStop}%
\bibitem [{\citenamefont {Philippov}\ \emph {et~al.}(2015)\citenamefont
  {Philippov}, \citenamefont {Spitkovsky},\ and\ \citenamefont
  {Cerutti}}]{Philippov:2014mqa}%
  \BibitemOpen
  \bibfield  {author} {\bibinfo {author} {\bibfnamefont {A.~A.}\ \bibnamefont
  {Philippov}}, \bibinfo {author} {\bibfnamefont {A.}~\bibnamefont
  {Spitkovsky}}, \ and\ \bibinfo {author} {\bibfnamefont {B.}~\bibnamefont
  {Cerutti}},\ }\href {\doibase 10.1088/2041-8205/801/1/L19} {\bibfield
  {journal} {\bibinfo  {journal} {Astrophys. J.}\ }\textbf {\bibinfo {volume}
  {801}},\ \bibinfo {pages} {L19} (\bibinfo {year} {2015})},\ \Eprint
  {http://arxiv.org/abs/1412.0673} {arXiv:1412.0673 [astro-ph.HE]} \BibitemShut
  {NoStop}%
\bibitem [{\citenamefont {Cerutti}\ and\ \citenamefont
  {Beloborodov}(2017)}]{Cerutti:2016ttn}%
  \BibitemOpen
  \bibfield  {author} {\bibinfo {author} {\bibfnamefont {B.}~\bibnamefont
  {Cerutti}}\ and\ \bibinfo {author} {\bibfnamefont {A.}~\bibnamefont
  {Beloborodov}},\ }\href {\doibase 10.1007/s11214-016-0315-7} {\bibfield
  {journal} {\bibinfo  {journal} {Space Sci. Rev.}\ }\textbf {\bibinfo {volume}
  {207}},\ \bibinfo {pages} {111} (\bibinfo {year} {2017})},\ \Eprint
  {http://arxiv.org/abs/1611.04331} {arXiv:1611.04331 [astro-ph.HE]}
  \BibitemShut {NoStop}%
\bibitem [{\citenamefont {Kalapotharakos}\ \emph {et~al.}(2018)\citenamefont
  {Kalapotharakos}, \citenamefont {Brambilla}, \citenamefont {Timokhin},
  \citenamefont {Harding},\ and\ \citenamefont
  {Kazanas}}]{Kalapotharakos:2017bpx}%
  \BibitemOpen
  \bibfield  {author} {\bibinfo {author} {\bibfnamefont {C.}~\bibnamefont
  {Kalapotharakos}}, \bibinfo {author} {\bibfnamefont {G.}~\bibnamefont
  {Brambilla}}, \bibinfo {author} {\bibfnamefont {A.}~\bibnamefont {Timokhin}},
  \bibinfo {author} {\bibfnamefont {A.~K.}\ \bibnamefont {Harding}}, \ and\
  \bibinfo {author} {\bibfnamefont {D.}~\bibnamefont {Kazanas}},\ }\href
  {\doibase 10.3847/1538-4357/aab550} {\bibfield  {journal} {\bibinfo
  {journal} {Astrophys. J.}\ }\textbf {\bibinfo {volume} {857}},\ \bibinfo
  {pages} {44} (\bibinfo {year} {2018})},\ \Eprint
  {http://arxiv.org/abs/1710.03170} {arXiv:1710.03170 [astro-ph.HE]}
  \BibitemShut {NoStop}%
\bibitem [{\citenamefont {{Ravenhall}}\ and\ \citenamefont
  {{Pethick}}(1994)}]{1994ApJ...424..846R}%
  \BibitemOpen
  \bibfield  {author} {\bibinfo {author} {\bibfnamefont {D.~G.}\ \bibnamefont
  {{Ravenhall}}}\ and\ \bibinfo {author} {\bibfnamefont {C.~J.}\ \bibnamefont
  {{Pethick}}},\ }\href {\doibase 10.1086/173935} {\bibfield  {journal}
  {\bibinfo  {journal} {\apj}\ }\textbf {\bibinfo {volume} {424}},\ \bibinfo
  {pages} {846} (\bibinfo {year} {1994})}\BibitemShut {NoStop}%
\bibitem [{\citenamefont {Kaplan}\ and\ \citenamefont {van
  Kerkwijk}(2009)}]{Kaplan:2009ce}%
  \BibitemOpen
  \bibfield  {author} {\bibinfo {author} {\bibfnamefont {D.~L.}\ \bibnamefont
  {Kaplan}}\ and\ \bibinfo {author} {\bibfnamefont {M.~H.}\ \bibnamefont {van
  Kerkwijk}},\ }\href {\doibase 10.1088/0004-637X/705/1/798} {\bibfield
  {journal} {\bibinfo  {journal} {Astrophys. J.}\ }\textbf {\bibinfo {volume}
  {705}},\ \bibinfo {pages} {798} (\bibinfo {year} {2009})},\ \Eprint
  {http://arxiv.org/abs/0909.5218} {arXiv:0909.5218 [astro-ph.HE]} \BibitemShut
  {NoStop}%
\bibitem [{\citenamefont {Dewdney}()}]{SKA}%
  \BibitemOpen
  \bibfield  {author} {\bibinfo {author} {\bibfnamefont {P.}~\bibnamefont
  {Dewdney}},\ }\href@noop {} {\enquote {\bibinfo {title} {{SKA1 SYSTEM
  BASELINE DESIGN}},}\ }\bibinfo {howpublished}
  {{\url{https://www.skatelescope.org/wp-content/uploads/2012/07/SKA-TEL-SKO-DD-001-1_BaselineDesign1.pdf}}},\
  \bibinfo {note} {accessed: 2019-10-29}\BibitemShut {NoStop}%
\bibitem [{\citenamefont {Drukier}\ \emph {et~al.}(1986)\citenamefont
  {Drukier}, \citenamefont {Freese},\ and\ \citenamefont
  {Spergel}}]{PhysRevD.33.3495}%
  \BibitemOpen
  \bibfield  {author} {\bibinfo {author} {\bibfnamefont {A.~K.}\ \bibnamefont
  {Drukier}}, \bibinfo {author} {\bibfnamefont {K.}~\bibnamefont {Freese}}, \
  and\ \bibinfo {author} {\bibfnamefont {D.~N.}\ \bibnamefont {Spergel}},\
  }\href {\doibase 10.1103/PhysRevD.33.3495} {\bibfield  {journal} {\bibinfo
  {journal} {Phys. Rev. D}\ }\textbf {\bibinfo {volume} {33}},\ \bibinfo
  {pages} {3495} (\bibinfo {year} {1986})}\BibitemShut {NoStop}%
\bibitem [{\citenamefont {Oliphant}(06  )}]{numpy}%
  \BibitemOpen
  \bibfield  {author} {\bibinfo {author} {\bibfnamefont {T.}~\bibnamefont
  {Oliphant}},\ }\href {http://www.numpy.org/} {\enquote {\bibinfo {title}
  {{NumPy}: A guide to {NumPy}},}\ }\bibinfo {howpublished} {USA: Trelgol
  Publishing} (\bibinfo {year} {2006--}),\ \bibinfo {note} {[Online; accessed
  08/05/2019]}\BibitemShut {NoStop}%
\bibitem [{\citenamefont {Jones}\ \emph {et~al.}(01  )\citenamefont {Jones},
  \citenamefont {Oliphant}, \citenamefont {Peterson} \emph {et~al.}}]{scipy}%
  \BibitemOpen
  \bibfield  {author} {\bibinfo {author} {\bibfnamefont {E.}~\bibnamefont
  {Jones}}, \bibinfo {author} {\bibfnamefont {T.}~\bibnamefont {Oliphant}},
  \bibinfo {author} {\bibfnamefont {P.}~\bibnamefont {Peterson}},  \emph
  {et~al.},\ }\href {http://www.scipy.org/} {\enquote {\bibinfo {title}
  {{SciPy}: Open source scientific tools for {Python}},}\ } (\bibinfo {year}
  {2001--}),\ \bibinfo {note} {[Online; accessed 08/05/2019]}\BibitemShut
  {NoStop}%
\end{thebibliography}%
